\documentclass[aps,prd,twocolumn,showpacs,floatfix,superscriptaddress,preprintnumbers]{revtex4}

\usepackage{amssymb}

\usepackage{epsfig}
\usepackage{graphicx}
\usepackage{stmaryrd}

\usepackage{rotating}
\usepackage[dvips]{color}
\definecolor{Black}{named}{Black}
\definecolor{Red}{named}{Red}

\def\lsim{\raise0.3ex\hbox{$\;<$\kern-0.75em\raise-1.1ex\hbox{$\sim\;$}}}
\def\gsim{\raise0.3ex\hbox{$\;>$\kern-0.75em\raise-1.1ex\hbox{$\sim\;$}}}
\renewcommand{\baselinestretch}{1.3}

\def\theta{\vartheta}
\def\phi{\varphi}

\newcommand{\be}{\begin{equation}}
\newcommand{\ee}{\end{equation}}
\newcommand{\bea}{\begin{eqnarray}}
\newcommand{\eea}{\end{eqnarray}}

\begin{document}

\preprint{IFIC/07-03}
%%{\noindent  IFIC/07-03}\\

\author{A.~Esteban-Pretel, R. Tom\`as and J.~W.~F.~Valle}
\affiliation{AHEP Group, Institut de F\'{\i}sica Corpuscular -
  C.S.I.C/Universitat de Val\`encia\\
Edifici Instituts d'Investigaci\'o, Apt. 22085, E-46071 Val\`encia, Spain}

\title{Probing non-standard neutrino interactions with supernova
  neutrinos}

\date{\today}

\begin{abstract}
  We analyze the possibility of probing non-standard neutrino
  interactions (NSI, for short) through the detection of neutrinos
  produced in a future galactic supernova (SN).
  We consider the effect of NSI on the neutrino propagation through
  the SN envelope within a three-neutrino framework,
  paying special attention to the inclusion of NSI-induced resonant
  conversions, which may take place in the most deleptonised inner
  layers.
  We study the possibility of detecting NSI effects in a Megaton water
  Cherenkov detector, either through modulation effects in the
  $\bar\nu_e$ spectrum due to (i) the passage of shock waves through
  the SN envelope, (ii) the time dependence of the electron fraction
  and (iii) the Earth matter effects; or, finally, through the
  possible detectability of the neutronization $\nu_e$ burst.
  We find that the $\bar\nu_e$ spectrum can exhibit dramatic features
  due to the internal NSI-induced resonant conversion. This occurs for
  non-universal NSI strengths of a few \%, and for very small
  flavor-changing NSI above a few$\times 10^{-5}$.
\end{abstract}


\maketitle

%%%%%%%%%%%%%%%%%%%%%%%%%%%%%%%%%%%%%%%%%%%%%%%%%%%%%%%%%%%%%%%%%%%%%%
\section{Introduction}
\label{sec:introduction}
%%%%%%%%%%%%%%%%%%%%%%%%%%%%%%%%%%%%%%%%%%%%%%%%%%%%%%%%%%%%%%%%%%%%%%

The very first data of the KamLAND collaboration~\cite{Eguchi:2002dm}
have been enough to isolate neutrino oscillations as the correct
mechanism explaining the solar neutrino
problem~\cite{Pakvasa:2003zv,Barger:2003qi}, indicating also that
large mixing angle (LMA) was the right solution.  The 766.3 ton-yr
KamLAND data sample further strengthens the validity of the LMA
oscillation interpretation of the data~\cite{Araki:2004mb}. 

Current data imply that neutrino have mass.  For an updated review of
the current status of neutrino oscillations see~\cite{Maltoni:2004ei}.
Theories of neutrino mass~\cite{schechter:1980gr,Valle:2006vb}
typically require that neutrinos have non-standard properties such as
neutrino electromagnetic transition
moments~\cite{Schechter:1981hw,Lim:1987tk,Akhmedov:1988uk} or
non-standard four-Fermi interactions (NSI, for
short)~\cite{Wolfenstein:1977ue,MS,Valle:1987gv}. The expected
magnitude of the NSI effects is rather model-dependent.

Seesaw-type models lead to a non-trivial structure of the lepton
mixing matrix characterizing the charged and neutral current weak
interactions~\cite{schechter:1980gr}. The NSI which are induced by the
charged and neutral current gauge interactions may be
sizeable~\cite{Mohapatra:1986bd,Bernabeu:1987gr,Branco:1989bn,Rius:1989gk,Deppisch:2004fa}.
Alternatively, non-standard neutrino interactions may arise in models
where neutrinos masses are radiatively
``calculable''~\cite{Zee:1980ai,Babu:1988ki}. Finally, in some
supersymmetric unified models, the strength of non-standard neutrino
interactions may arise from renormalization and/or threshold
effects~\cite{Hall:1985dx}.

We stress that non-standard interactions strengths are highly
model-dependent.  In some models NSI strengths are too small to be
relevant for neutrino propagation, because they are either suppressed
by some large mass scale or restricted by limits on neutrino masses,
or both.  However, this need not be the case, and there are many
theoretically attractive scenarios where moderately large NSI
strengths are possible and consistent with the smallness of neutrino
masses. In fact one can show that NSI may exist even in the limit of
massless
neutrinos~\cite{Mohapatra:1986bd,Bernabeu:1987gr,Branco:1989bn,Rius:1989gk,Deppisch:2004fa}.
Such may also occur in the context of fully unified models like
$SO(10)$~\cite{Malinsky:2005bi}.

We argue that, in addition to the precision determination of the
oscillation parameters, it is necessary to test for sub-leading
non-oscillation effects that could arise from non-standard neutrino
interactions. These are natural outcome of many neutrino mass models
and can be of two types: flavor-changing (FC) and non-universal (NU).
These are constrained by existing experiments (see below) and, with
neutrino experiments now entering a precision
phase~\cite{McDonald:2004dd}, an improved determination of neutrino
parameters and their theoretical impact constitute an important goal
in astroparticle and high energy physics~\cite{Maltoni:2004ei}.

%%%% motivation for using SN neutrinos to study NSI

Here we concentrate on the impact of non-standard neutrino
interactions on supernova physics. We show how complementary
information on the NSI parameters  could be inferred from the
detection of core-collapse supernova neutrinos.
The motivation for the study is twofold. First, if a future SN event
takes place in our Galaxy the number of neutrino events expected in
the current or planned neutrino detectors would be enormous,
$\mathcal{O}(10^4-10^5)$~\cite{Scholberg:2007nu}.  Moreover, the
extreme conditions under 
which neutrinos have to travel since they are created in the SN core,
in strongly deleptonised regions at nuclear densities, until they
reach the Earth, lead to strong matter effects. In particular the
effect of small values of the NSI parameters can be dramatically
enhanced, possibly leading to observable consequences.

This paper is planned as follows.  In Sec.~\ref{sec:preliminaries} we
summarize the current observational bounds on the parameters
describing the NSI, including previous works on NSI in SNe. In
Sec.~\ref{sec:neutrino-evolution} we describe the neutrino propagation
formalism as well as the SN profiles which will be used. In
Sec.~\ref{sec:two-regimes} we analyze the effect of NSI on the $\nu$
propagation in the inner regions near the neutrinosphere and in the
outer regions of the SN envelope. In Sec.~\ref{sec:observables} we
discuss the possibility of using various observables to probe the
presence of NSI in the neutrino signal of a future galactic SN.
Finally in Sec.~\ref{sec:summary} we present our conclusions.

\section{Preliminaries}
\label{sec:preliminaries}

A large class of non-standard interactions may be parametrized with
the effective low-energy four-fermion operator:
\begin{equation}
\label{eq:Lnsi}
\mathcal{L}_{NSI} = -\varepsilon^{fP}_{\alpha\beta}
2\sqrt{2}G_F(\bar\nu_\alpha\gamma_\mu L \nu_\beta) (\bar f\gamma^\mu P f)~,
\end{equation}
where $P=L,~R$ and $f$ is a first generation fermion: $e,~u,~d$. The
coefficients $\varepsilon^{fP}_{\alpha\beta}$ denote the strength of
the NSI between the neutrinos of flavors $\alpha$ and $\beta$ and the
$P-$handed component of the fermion $f$.

%%%%%%%%%% current limits 

Current constraints on $\varepsilon^{fP}_{\alpha\beta}$ come from a
variety of different sources, which we now briefly list.

\subsection{Laboratory}
\label{sec:laboratory}

Neutrino scattering
experiments~\cite{Auerbach:2001wg,Daraktchieva:2003dr,Dorenbosch:1986tb,Vilain:1994qy,Zeller:2001hh}
provide the following bounds, $|\varepsilon^{fP}_{\mu\mu}| \lesssim
10^{-3}-10^{-2},~ |\varepsilon^{fP}_{ee}|\lesssim
10^{-1}-1,~|\varepsilon^{fP}_{\mu\tau}|\lesssim 0.05,~
|\varepsilon^{fP}_{e\tau}|\lesssim 0.5$ at 90 \%
C.L~\cite{Barger:1991ae,Davidson:2003ha,Barranco:2005ps}.
On the other hand the analysis of the $e^+e^-\to \nu\bar\nu\gamma$
cross section measured at LEP II  leads to a
bound on $|\varepsilon^{eP}_{\tau\tau}|\lesssim
0.5$~\cite{Berezhiani:2001rs}.
Future prospects to improve the current limits imply the measurement
of $\sin^2\theta_W$ leptonically in the scattering off electrons in
the target, as well as in neutrino deep inelastic scattering in a
future neutrino factory. The main improvement would be in the case of
$|\varepsilon^{fP}_{ee}|$ and $|\varepsilon^{fP}_{e\tau}|$, where
values as small as $10^{-3}$ and $0.02$, respectively, could be
reached~\cite{Davidson:2003ha}.

The search for flavor violating processes involving charged leptons is
expected to restrict corresponding neutrino interactions, to the
extent that the $SU(2)$ gauge symmetry is assumed. However, this can
at most give indicative order-of-magnitude restrictions, since we know
$SU(2)$ is not a good symmetry of nature. 
Using radiative corrections it has been argued that, for example,
$\mu-e$ conversion on nuclei like in the case of $\mu^-Ti$ also
constrains $|\varepsilon^{qP}_{\mu e}|\lesssim 7.7\times
10^{-4}$~\cite{Davidson:2003ha}.

Non-standard interactions can also affect neutrino propagation through
matter, probed in current neutrino oscillation experiments.  The
bounds so obtained apply to the vector coupling constant of the NSI,
$\varepsilon^{fV}_{\alpha\beta} = \varepsilon^{fL}_{\alpha\beta} +
\varepsilon^{fR}_{\alpha\beta}$, since only this appears in neutrino
propagation in matter~\footnote{ Axial couplings would affect neutrino
  propagation in polarized media, see Ref.~\cite{Nunokawa:1997dp}.}.

\subsection{Solar and reactor}
\label{sec:solar-reactor}

The role of neutrino NSIs as subleading effects on the solar
neutrino oscillations and KamLAND has been recently considered in
Ref.~\cite{Friedland:2004pp,Guzzo:2004ue,Miranda:2004nb} with the
following bounds at 90 \% CL for $\varepsilon \equiv -\sin \theta_{23}
\varepsilon^{dV}_{e\tau}$ with the allowed range $-0.93\lesssim
\varepsilon \lesssim 0.30$, while for the diagonal term
$\varepsilon'\equiv
\sin^2\theta_{23}\varepsilon^{dV}_{\tau\tau}-\varepsilon^{dV}_{ee}$,
the only forbidden region is $[0.20,0.78]$~\cite{Miranda:2004nb}.
Only in the ideal case of infinitely precise solar neutrino
oscillation parameters determination, the allowed range would ``close
from the left'' for negative NSI parameter values, at $-0.6$ for
$\varepsilon$ and $-0.7$ for $\varepsilon'$.

\subsection{Atmospheric and accelerator neutrinos}
\label{sec:atmosph-accel-neutr}

Non-standard interactions involving muon neutrinos can be constrained by
atmospheric neutrino experiments as well as accelerator neutrino
oscillation searches at K2K and MINOS.  In Ref.~\cite{Fornengo:2001pm}
Super-Kamiokande and MACRO observations of atmospheric neutrinos were
considered in the framework of two neutrinos.  The limits obtained
were $-0.05\lesssim \varepsilon^{dV}_{\mu\tau}<0.04$ and
$|\varepsilon^{dV}_{\tau\tau}-\varepsilon^{dV}_{\mu\mu}|\lesssim 0.17$
at 99 \% CL. The same data set together with K2K were recently
considered in Refs.~\cite{Friedland:2004ah,Friedland:2005vy} to study
the nonstandard neutrino interactions in a three generation scheme
under the assumption
$\varepsilon_{e\mu}=\varepsilon_{\mu\mu}=\varepsilon_{\mu\tau}=0$. 
The allowed region of $\varepsilon_{\tau\tau}$ obtained 
for values of $\varepsilon_{e\tau}$ smaller than
$\mathcal{O}(10^{-1})$ becomes
$\Sigma_{f=u,d,e}\varepsilon^{fV}_{\alpha\beta}N_f/N_e \lesssim
0.2$~\cite{Friedland:2005vy} , where $N_f$ stands for the fermion
number density.

\subsection{Cosmology}
\label{sec:cosmology}

If non-standard interactions with electrons were large they might also
lead to important cosmological and astrophysical implications. For
instance, neutrinos could be kept in thermal contact with electrons
and positrons longer than in the standard case, hence they would share
a larger fraction of the entropy release from $e^\pm$ annihilations.
This would affect the predicted features of the cosmic background of
neutrinos.  As recently pointed out in Ref~\cite{Mangano:2006ar}
required couplings are, though, larger than the current laboratory
bounds.
%

%%%%%%%%%%%%%%%%%%%%%%%%%%%%%%%%%%%%%%%%%%%%%%%%%%%%%%%%%%%%%%%%%%%%%%
\subsection{NSI in Supernovae}
\label{sec:nsi_sn}
%%%%%%%%%%%%%%%%%%%%%%%%%%%%%%%%%%%%%%%%%%%%%%%%%%%%%%%%%%%%%%%%%%%%%%

According to the currently accepted supernova (SN) paradigm, neutrinos
are expected to play a crucial role in SN dynamics. As a result, SN
physics provides a laboratory to probe neutrino properties.  Moreover,
many future large neutrino detectors are currently being
discussed~\cite{Katsanevas:2006}.  The enormous number of events,
$\mathcal{O}(10^4-10^5)$ that would be ``seen'' in these detectors
indicates that a future SN in our Galaxy would provide a very
sensitive probe of non-standard neutrino interaction effects.

The presence of NSI can lead to important consequences for the SN
neutrino physics both in the highly dense core as well as in the
envelope where neutrinos basically freely stream.

The role of non-forward neutrino scattering processes on heavy nuclei
and free nucleons giving rise to flavor change within the SN core has
been recently analyzed in Ref.~\cite{Amanik:2004vm,Amanik:2006ad}. The
main effect found was a reduction in the core electron fraction $Y_e$
during core collapse. A lower $Y_e$ would lead to a lower homologous
core mass, a lower shock energy, and a greater nuclear
photon-disintegration burden for the shock wave. By allowing a maximum
$\Delta Y_e = -0.02$ it has been claimed that
$\varepsilon_{e\alpha}\lesssim 10^{-3}$, where
$\alpha=\mu,~\tau$~\cite{Amanik:2006ad}.

On the other hand it has been noted since long ago that the existence
of NSI plays an important role in the propagation of SN neutrinos
through the envelope leading to the possibility of a new resonant
conversion. In contrast to the well known MSW
effect~\cite{Mikheev:1986gs,Mikheev:1986wj} it would take place even
for massless neutrinos~\cite{Valle:1987gv}. Two basic ingredients are
necessary: universal and flavor changing NSI.  In the original scheme
neutrinos were mixed in the leptonic charged current and universality
was violated thanks to the effect of mixing with heavy gauge singlet
leptons~\cite{schechter:1980gr,Mohapatra:1986bd}.
Such resonance would induce strong neutrino flavor conversion both for
neutrinos and antineutrinos simultaneously, possibly affecting the
neutrino signal of the SN1987A as well as the possibility of having
$r-$process nucleosynthesis.
This was first quantitatively considered within a two-flavor
$\nu_e-\nu_\tau$ scheme, and bounds on the relevant NSI parameters
were obtained using both arguments~\cite{Nunokawa:1996tg}.

One of the main features of the such ``internal'' or ``massless''
resonant conversion mechanism is that it requires the violation of
universality, its position being determined only by the matter
chemical composition, namely the value of the electron fraction $Y_e$,
and not by the density.
In view of the experimental upper bounds on the NSI parameters such
new resonance can only take place in the inner layers of the
supernova, near the neutrinosphere, where $Y_e$ takes its minimum
values.  In this region the values of $Y_e$ are small enough to allow
for resonance conversions to take place in agreement with existing
bounds on the strengths of non-universal NSI parameters.

The SN physics implications of another type of NSI present in
supersymmetric R-parity violating models have also been studied in
Ref.~\cite{Nunokawa:1996ve}, again for a system of two neutrinos. For
definiteness NSI on $d-$quarks were considered, in two cases: (i)
massless neutrinos without mixing in the presence of flavor-changing
(FC) and non-universal (NU) NSIs, and (ii) neutrinos with eV masses
and FC NSI.
Different arguments have been used in order to constrain the parameters
describing the NSI, namely, the SN1987A signal, the possibility to get
successful $r-$process nucleosynthesis, and the possible enhancement
of the energy deposition behind the shock wave to reactivate it.

On the other hand several subsequent
articles~\cite{Mansour:1997fi,Bergmann:1998rg,Fogli:2002xj} considered
the effects of NSI on the neutrino propagation in a three--neutrino
mixing scenario for the case $Y_e >0.4$, typical for the outer SN
envelope.  Together with the assumption that
$\varepsilon^{dV}_{\alpha\beta}\lesssim 10^{-2}$ this prevents the
appearance of internal resonances in contrast to previous references.

Motivated by supersymmetric theories without R parity, in
Ref.~\cite{Mansour:1997fi} the authors considered the effects of
small-strength NSI with $d-$quarks.  Following the formalism developed
in Refs.~\cite{Kuo:1987qu,Bergmann:1997mr} they studied the
corrections that such NSI would have on the expressions for the
survival probabilities in the standard resonances MSW-H and MSW-L.
A similar analysis was performed in Ref.~\cite{Bergmann:1998rg}
assuming Z-induced NSI interactions originated by additional heavy
neutrinos.
A phenomenological generalization of these results was carried out in
Ref.~\cite{Fogli:2002xj}. The authors found an analytical compact
expression for the survival probabilities in which the main effects of
the NSI can be embedded through shifts of the mixing angles
$\theta_{12}$ and $\theta_{13}$. In contrast to similar expressions
found previously these directly apply to all mixing angles, and in the
case with Earth matter effects. The main phenomenological consequence
was the identification of a degeneracy between $\theta_{13}$ and
$\varepsilon_{e\alpha}$, similar to the analogous ``confusion''
between $\theta_{13}$ and the corresponding NSI parameter noted to
exist in the context of long-baseline neutrino
oscillations~\cite{huber:2001de,huber:2002bi}.

We have now re-considered the general three--neutrino mixing scenario
with NSI. In contrast to previous
work~\cite{Mansour:1997fi,Bergmann:1998rg,Fogli:2002xj}, we have not
restricted ourselves to large values of $Y_e$, discussing also small
values present in the inner layers.  This way our generalized
description includes both the possibility of neutrinos having the
``massless'' NSI-induced resonant conversions in the inner layers of
the SN envelope~\cite{Valle:1987gv,Nunokawa:1996tg,Nunokawa:1996ve},
as well as the ``outer'' oscillation-induced
conversions~\cite{Mansour:1997fi,Bergmann:1998rg,Fogli:2002xj}~\footnote{However
  we have confined ourselves to values of $\varepsilon_{e\alpha}$
  small enough not to lead to drastic consequences during the core
  collapse.}.

%%%%%%%%%%%%%%%%%%%%%%%%%%%%%%%%%%%%%%%%%%%%%%%%%%%%%%%%%%%%%%%%%%%%%%

\section{Neutrino evolution}
\label{sec:neutrino-evolution}

%%%%%%%%%%%%%%%%%%%%%%%%%%%%%%%%%%%%%%%%%%%%%%%%%%%%%%%%%%%%%%%%%%%%%%

In this section we describe the main ingredients of our analysis.  Our
emphasis will be on the use of astrophysically realistic SN matter and
$Y_e$ profiles, characterizing its density and the matter composition.
Their details, in particular their time dependence, are crucial in
determining the way the non-standard 
neutrino interactions affect the propagation of neutrinos in the SN
medium.

\subsection{Evolution Equation}
\label{sec:evolution-equation}

As discussed in Sec.~\ref{sec:preliminaries} in an unpolarized medium
the neutrino propagation in matter will be affected by the vector
coupling constant of the NSI, $\varepsilon_{\alpha\beta}^{fV}=
\varepsilon_{\alpha\beta}^{fL} +
\varepsilon_{\alpha\beta}^{fR}$~\footnote{For the sake of simplicity 
  we will omit the superindex $V$.}. 
The way the neutral current NSI modifies the neutrino evolution will
be parametrized phenomenologically through the effective low-energy
four-fermion operator described in Eq.~(\ref{eq:Lnsi}).
We also assume $\varepsilon^f_{\alpha\beta} \in\Re$, neglecting
possible $CP$ violation in the new interactions.

Under these assumptions the Hamiltonian describing the SN neutrino
evolution in the presence of NSI can be cast in the following
form~\footnote{The importance of collective flavor neutrino
  conversions driven by neutrino-neutrino interactions has been
  recently noted in
  Refs.~\cite{Duan:2006an,Duan:2006jv,Hannestad:2006nj,Raffelt:2007yz}. Here we
  consider only the case where the effective potential felt by
  neutrinos comes from their interactions with electrons, protons and
  neutrons. In a future work we plan to include this effect and have a
  complete picture of the neutrino propagation.}
\begin{equation}
{\rm i}\frac{{\rm d}}{{\rm d}r}
\nu_\alpha
%\left(\begin{array}{c}\nu_e \\ \nu_\mu
%  \\ \nu_\tau \end{array} \right)
=
\left(
H_{\rm kin} + H_{\rm int} 
\right)_{\alpha\beta}
\nu_\beta
% \left(\begin{array}{c}\nu_e \\ \nu_\mu
%  \\ \nu_\tau \end{array} \right)
~,
\end{equation}
where $H_{\rm kin}$ stands for the kinetic term
\begin{equation}
H_{\rm kin} = 
U\frac{M^2}{2E}U^\dagger~, 
%U\frac{1}{2E} \left( \begin{array}{ccc} m_1^2 & 0 & 0 \\ 0
%  & m_2^2 & 0 \\ 0 & 0 & m_3^2 \end{array} \right)
%U^\dagger~,
\end{equation}
with $M^2={\rm diag}(m_1^2,m_2^2,m_3^2)$, and $U$ the three-neutrino
lepton mixing matrix~\cite{schechter:1980gr} in the PDG
convention~\cite{Yao:2006px} and with no $CP$ phases.

The second term of the Hamiltonian accounts for the interaction of
neutrinos with matter and can be split into two pieces,
\begin{equation}
H_{\rm int} = H^{\rm std}_{\rm int} + H^{\rm nsi}_{\rm int}~.
\end{equation}
The first term, $ H^{\rm std}_{\rm int}$ describes the standard
interaction with matter and can be written as $H^{\rm std}_{\rm int}$
= diag $(V_{CC},0,0)$ up to one loop corrections due to different
masses of the muon and tau leptons~\cite{Botella:1986wy}.
The standard matter potential for neutrinos is given by
\begin{eqnarray}
V_{CC} & = & \sqrt{2}G_F N_e = V_0 \rho Y_e~,
\end{eqnarray}
where $V_0\approx 7.6\times 10^{-14}$~eV, the density is given in
${\rm g/cm}^3$, and $Y_e$ stands for the relative number of electrons
with respect to baryons.  For antineutrinos the potential is identical
but with the sign changed.

The term in the Hamiltonian describing the non-standard neutrino
interactions with a fermion $f$ can be expressed as,
\begin{equation}
(H_{\rm int}^{\rm nsi})_{\alpha\beta} = \sum_{f=e,u,d} (V_{\rm
    nsi}^{f})_{\alpha\beta}~,
\end{equation}
with $(V_{\rm nsi}^f)_{\alpha\beta} \equiv \sqrt{2}G_F
N_f\varepsilon^f_{\alpha\beta}$.
For definiteness and motivated by actual models, for example, those
with broken R parity supersymmetry we take for $f$ the down-type
quark. However, an analogous treatment would apply to the case of NSI
on up-type quarks, the existence of NSI with electrons brings no
drastic qualitative differences with respect to the pure oscillation
case (see below).  Therefore the NSI potential can be expressed as
follows,
\begin{equation}
(V_{\rm nsi}^{d})_{\alpha\beta} 
=\varepsilon^{d}_{\alpha\beta}V_0\rho(2-Y_e)~. 
\end{equation}
From now on we will not explicitely write the superindex $d$.  In
order to further simplify the problem we will redefine the diagonal
NSI parameters so that $\varepsilon_{\mu\mu}=0$, as one can easily see
that subtracting a matrix proportional to the identity leaves the
physics involved in the neutrino oscillation unaffected.

\subsection{Supernova matter profiles}
\label{sec:supern-matt-prof}

\begin{figure}[h]
  \begin{center}
    \includegraphics[width=0.5\textwidth]{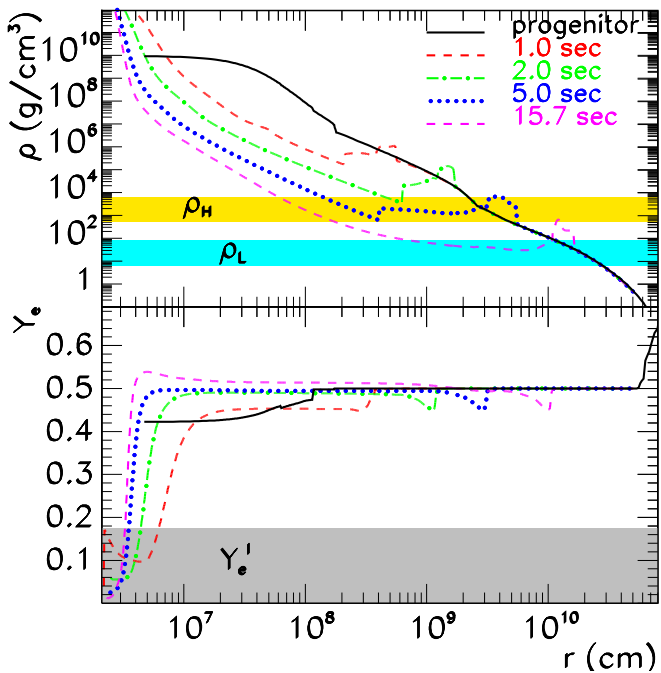}
  \end{center}
  \caption{Density (upper panel) and electron fraction (bottom panel)
    profiles for the SN progenitor and at different instants after the
    core bounce, from Ref.~\cite{Tomas:2004gr}. The regions where the
    $H$ (yellow) and the $L$ (cyan) resonance take place are also
    indicated, as well as the NSI-induced $I$ (gray) resonance for the
    parameters $\varepsilon_{ee}=0,~\varepsilon_{\tau\tau}\lesssim
    0.07$ and $|\varepsilon_{\mu\tau}|\lesssim 0.05$ }
  \label{fig:snprofiles}
\end{figure}

Neutrino propagation depends on the supernova matter and chemical
profile through the effective potential.  This profile exhibits an
important time dependence during the explosion.
Fig.~\ref{fig:snprofiles} shows the density $\rho(t,r)$ and the
electron fraction $Y_e(t,r)$ profiles for the SN progenitor as well as
at different times post-bounce.

Progenitor density profiles can be roughly parametrized by a power-law
function 
\begin{equation}
\rho(r) = \rho_0 \left(\frac{R_0}{r}\right)^n~,
\end{equation}
where $\rho_0 \sim 10^4$~g/cm$^3$, $R_0\sim 10^9$~cm, and $n\sim 3$.
The electron fraction profile varies depending on the matter
composition of the different layers.  For instance, typical values of
$Y_e$ between 0.42 and 0.45 in the inner regions are found in stellar
evolution simulations~\cite{2002RvMP...74.1015W}. In the intermediate
regions, where the MSW $H$ and $L$-resonances take place $Y_e\approx
0.5$.  This value can further increase in the most outer layers of the
SN envelope due to the presence of hydrogen.

After the SN core bounce the matter profile is affected in several ways.
First note that a front shock wave starts to propagate outwards and
eventually ejects the SN envelope. The evolution of the shock wave
will strongly modify the density profile and therefore the neutrino
propagation~\cite{Schirato:2002tg,Fogli:2003dw}. Following
Ref.~\cite{Tomas:2004gr} we shall assume that the structure of the
shock wave is more complicated and an additional ``reverse wave''
appears due to the collision of the neutrino-driven wind and the
slowly moving material behind the forward shock, as seen in the upper
panel of Fig.~\ref{fig:snprofiles}~\footnote{Here we neglect the
  possible effects of density
  fluctuations~\cite{balantekin:1996pp,nunokawa:1996qu} taking place
  during the shock wave propagation. For a detailed study of the
  phenomenological consequences see
  Refs.~\cite{Fogli:2006xy,Friedland:2006ta}.}.

On the other hand, the electron fraction is also affected by the time
evolution as the SN explosion proceeds.
Once the collapse starts the core density grows so that the neutrinos
become eventually effectively trapped within the so called
``neutrinosphere''. At this point the trapped electron fraction has
decreased until values of the order of
0.33~\cite{Cardall:2007dy}. When the inner core reaches the nuclear
density it can not contract any further and bounces. As a consequence
a shock wave forms in the inner core and starts propagating outwards.
When the newly formed supernova shock reaches densities low enough for
the initially trapped neutrinos to begin streaming faster than the
shock propagates~\cite{Bethe:1980gq}, a breakout pulse of $\nu_e$ is
launched.
In the shock-heated matter, which is still rich of electrons and
completely disintegrated into free neutrons and protons, a large
number of $\nu_e$ are rapidly produced by electron captures on
protons. They follow the shock on its way out until they are released
in a very luminous flash, the breakout burst, at about the moment when
the shock penetrates the neutrinosphere and the neutrinos can
escape essentially unhindered.  As a consequence, the lepton number in
the layer around the neutrinosphere decreases strongly and the matter
neutronizes~\cite{burrows_mazurek}.  The value of $Y_e$ steadily
decreases in these layers until values of the order of
$\mathcal{O}(10^{-2})$.  Outside the neutrinosphere there is a steep
rise until $Y_e\approx 0.5$. This is a robust feature of the
neutrino-driven baryonic wind. Neutrino heating drives the wind mass
loss and causes $Y_e$ to rise within a few $10$~km from low to high
values, between 0.45 and 0.55~\cite{private}, see bottom panel of
Fig.~\ref{fig:snprofiles}.
Inspired in the numerical results of Ref.~\cite{Tomas:2004gr} we have
parametrized the behavior of the electron fraction near the
neutrinosphere phenomenologically as,
\begin{equation}
Y_e = a + b\arctan[(r-r_0)/r_s]~,
\label{eq:Ye}
\end{equation}
where $a\approx 0.23-0.26$ and $b\approx 0.16-0.20$. The parameters $r_0$ and
$r_s$  describe where the rise takes place and how steep it is,
respectively. As can be seen in Fig.~\ref{fig:snprofiles} both
decrease with time.

%%%%%%%%%%%%%%%%%%%%%%%%%%%%%%%%%%%%%%%%%%%%%%%%%%%%%%%%%%%%%%%%%%%%%%%%
\section{The two regimes}
\label{sec:two-regimes}

%%%%%%%%%%%%%%%%%%%%%%%%%%%%%%%%%%%%%%%%%%%%%%%%%%%%%%%%%%%%%%%%%%%%%%%%

In order to study the neutrino propagation through the SN envelope we
will split the problem into two different regions: the inner envelope,
defined by the condition $V_{CC}\gg \Delta m^2_{\rm atm}/(2E)$ with
$\Delta m^2_{\rm atm} \equiv m_3^2-m_2^2$, and
the outer one, where $\Delta m^2_{\rm atm}/(2E) \gtrsim V_{CC}$. From
the upper panel of Fig.~\ref{fig:snprofiles} one can see how the
boundary roughly varies between $r\approx 10^8$~cm and $10^9$~cm,
depending on the time considered.
This way one can fully characterize all resonances that can take place
in the propagation of supernova neutrinos, both the outer resonant
conversions related to neutrino masses and indicated as the upper
bands in Fig.~\ref{fig:snprofiles}, and the inner resonances that
follow from the presence of non-standard neutrino interactions,
indicated by the band at the bottom of the same figure.
Here we pay special attention to the use of realistic matter and
chemical supernova profiles and three-neutrino flavors thus
generalising previous studies.

%%%%%%%%%%%%%%%%%%%%%%%%%%%%%%%%%%%%%%%%%%%%%%%%%%%%%%%%%%%%%%%%%%%%%%%%
\subsection{Neutrino Evolution in the Inner Regions}
\label{subsec:nu_evolution_inner}
%%%%%%%%%%%%%%%%%%%%%%%%%%%%%%%%%%%%%%%%%%%%%%%%%%%%%%%%%%%%%%%%%%%%%%%%

Let us first write the Hamiltonian in the inner layers, where $H_{\rm
  int}\gg H_{\rm kin}$. In this case the Hamiltonian can be written as
\begin{equation}
H\approx H_{\rm int} =
V_0\rho(2-Y_e)
\left(\begin{array}{ccc} \frac{Y_e}{2-Y_e}+\varepsilon_{ee} &
  \varepsilon_{e\mu} & \varepsilon_{e\tau} \\
\varepsilon_{e\mu} & 0 & \varepsilon_{\mu\tau} \\
\varepsilon_{e\tau} & \varepsilon_{\mu\tau} & \varepsilon_{\tau\tau}  
\end{array} \right)~.
\end{equation}

When the value of the $\varepsilon_{\alpha\beta}$ is of the same order
as the electron fraction $Y_e$ internal resonances can
arise~\cite{Valle:1987gv}.
Taking into account the current constraints on the $\varepsilon$'s
discussed in Sec.~\ref{sec:preliminaries} one sees that small values
of $Y_e$ are required~\cite{Nunokawa:1996tg,Nunokawa:1996ve}.
As a result, these can only take place in the most deleptonised inner
layers, close to the neutrinosphere, where the kinetic terms of the
Hamiltonian are negligible.

Given the large number of free parameters $\varepsilon_{\alpha\beta}$
involved we consider one particular case where 
$|\varepsilon_{e\mu}|$ and $|\varepsilon_{e\tau}|$ are small enough to
neglect a possible initial mixing between $\nu_e$ and $\nu_\mu$ or
$\nu_\tau$. 
Barring fine tuning, this basically amounts to
$|\varepsilon_{e\mu}|,~|\varepsilon_{e\tau}|\ll 10^{-2}$. According to
the discussion of Sec.~\ref{sec:preliminaries} $\varepsilon_{e\mu}$
automatically satisfies the condition, whereas one expects that the
window $|\varepsilon_{e\tau}|\gtrsim 10^{-2}$ will eventually be
probed in future experiments.

Since the initial fluxes of $\nu_\mu$ and $\nu_\tau$ are expected to
be basically identical, it is convenient to redefine the weak basis by
performing a rotation in the $\mu-\tau$ sector:
\begin{equation}
\left(\begin{array}{c} \nu_e \\ \nu_\mu \\ \nu_\tau \end{array}\right)
= 
U(\theta_{23}') 
\left(\begin{array}{c} \nu_e \\ \nu_\mu' \\ \nu_\tau' \end{array}\right)
=
\left(\begin{array}{ccc} 1 & 0 & 0\\ 0 & c_{23'} &
  s_{23'} \\ 0 & -
  s_{23'} & c_{23'} 
\end{array}\right)
\left(\begin{array}{c} \nu_e \\ \nu_\mu' \\ \nu_\tau' \end{array}\right)~,
\end{equation}
where $c_{23'}$ and $s_{23'}$ stand for
$\cos(\theta_{23}')$ and
$\sin(\theta_{23}')$, respectively. 
The angle $\theta_{23}'$ can be written as
\begin{equation}
\tan(2\theta_{23}') \approx
\frac{2H_{23}}{H_{33}} =
  \frac{2\varepsilon_{\mu\tau}}{\varepsilon_{\tau\tau}}~.   
\end{equation}

The Hamiltonian becomes in the new basis
\begin{eqnarray}
H'_{\alpha\beta} & = &
U^\dagger(\theta_{23}')H_{\alpha\beta}
U(\theta_{23}') \\
& = &
V_0\rho(2-Y_e)
\left(\begin{array}{ccc}
\frac{Y_e}{2-Y_e}+\varepsilon_{ee} & \varepsilon_{e\mu}' &
\varepsilon_{e\tau}' \\
\varepsilon_{e\mu}' & \varepsilon_{\mu\mu}' & 0 \\
\varepsilon_{e\tau}' & 0 & \varepsilon_{\tau\tau}'
\end{array}\right)~,
\label{eq:HI1}
\end{eqnarray}
where
\begin{eqnarray}
\varepsilon_{e\mu}' & = & \varepsilon_{e\mu}c_{23'}-
  \varepsilon_{e\tau}s_{23'}\\
\varepsilon_{e\tau}' & = & \varepsilon_{e\mu}s_{23'}+
  \varepsilon_{e\tau}c_{23'}\\
\varepsilon_{\mu\mu}' & = &
  (\varepsilon_{\tau\tau}-
  \sqrt{\varepsilon^2_{\tau\tau}+4\varepsilon^2_{\mu\tau}})/2~\\    
\varepsilon_{\tau\tau}' & = &
  (\varepsilon_{\tau\tau}+
\sqrt{\varepsilon^2_{\tau\tau}+4\varepsilon^2_{\mu\tau}})/2~.
\end{eqnarray}

With our initial assumptions on $\varepsilon_{e\alpha}$ one notices
that the new basis $\nu_\alpha'$ basically diagonalizes the
Hamiltonian, and therefore coincides roughly with the matter
eigenstate basis.
A novel resonance can arise if the condition $H_{ee}'=H_{\tau\tau}'$
is satisfied, we call this $I$-resonance, $I$ standing for
``internal''~\footnote{The alternative condition $H_{ee}' =
  H_{\mu\mu}'$ would give rise to another internal resonance which can
  be studied using the same method. For brevity, we will not pursue
  this in this paper.}.  The corresponding resonance condition can be
written as
\begin{equation}
\label{eq:Ye_resI}
Y_e^{I} = \frac{2\varepsilon^{I}}{1+\varepsilon^{I}}~,
\end{equation}
where
$\varepsilon^{I}$ is defined as
$\varepsilon_{\tau\tau}'- \varepsilon_{ee}$. 
\begin{figure}
  \begin{center}
    \includegraphics[width=0.45\textwidth]{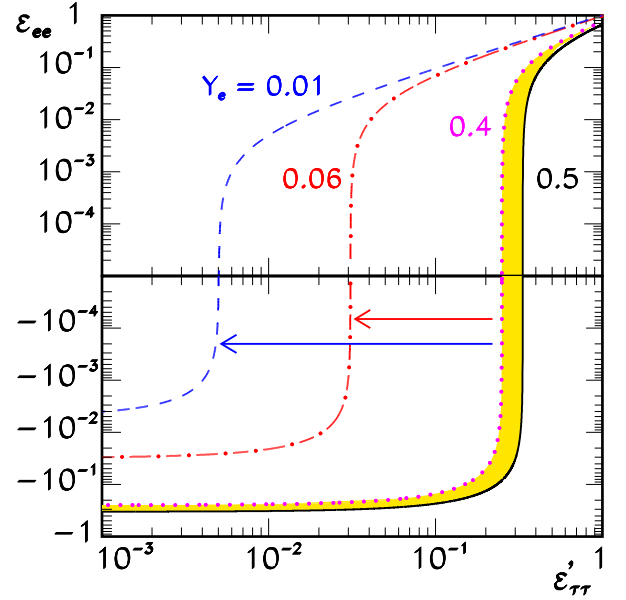}
  \end{center}
  \caption{Contours of $Y_e^{I}$ as function of $\varepsilon_{ee}$ and
    $\varepsilon_{\tau\tau}'$ according to Eq.~(\ref{eq:Ye_resI}) for
    different values of $Y_e$. The region in yellow represents the
    region of parameters that gives rise to $I$-resonance before the
    collapse. The arrows indicate how this region widens with time.}
  \label{fig:YeI-eI}
\end{figure}
In Fig.~\ref{fig:YeI-eI} we represent the range of $\varepsilon_{ee}$
and $\varepsilon_{\tau\tau}'$ leading to the $I$-resonance for an
electron fraction profile between different $Y_e^{\rm min}$'s and
$Y_e^{\rm max}=0.5$.  It is important to notice that the value of
$Y_e^{\rm min}$ depends on time. Right before the collapse the minimum
value of the electron fraction is around $0.4$. Hence the window of
NSI parameters that would lead to a resonance would be relatively
narrow, as indicated by the shaded (yellow) band in
Fig.~\ref{fig:YeI-eI}.
As time goes on $Y_e^{\rm min}$ decreases to values of the order of a
few \%, and as a result the region of parameters giving rise to the
$I$-resonance significantly widens.
For example, in the range $|\varepsilon_{ee}| \leq 10^{-3}$
possibly accessible to future experiments one sees that the
$I$-resonance can take place for values of $\varepsilon_{\tau\tau}'$
of the order of $\mathcal{O}(10^{-2})$. 
This indicates that the potential sensitivity on NSI parameters that
can be achieved in supernova studies is better than that of the
current limits. 
As seen in Fig.~\ref{fig:snprofiles} in order to fulfill the
$I$-resonance condition for such small values of the NSI parameters
the values of $Y_e$ must indeed lie, as already stated, in the inner
layers.

\begin{figure}
  \begin{center}
  \includegraphics[width=0.45\textwidth]{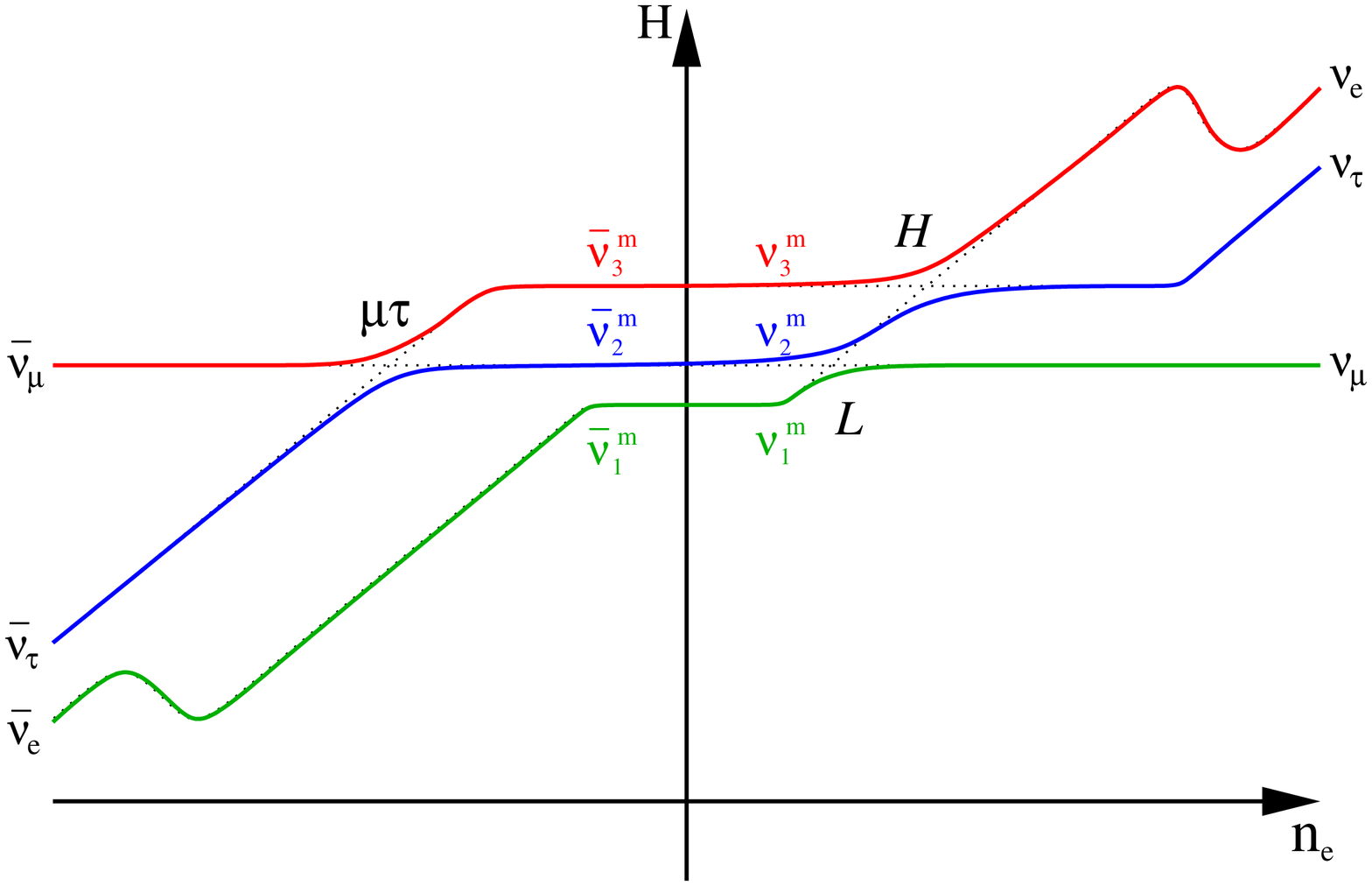}
  \includegraphics[width=0.45\textwidth]{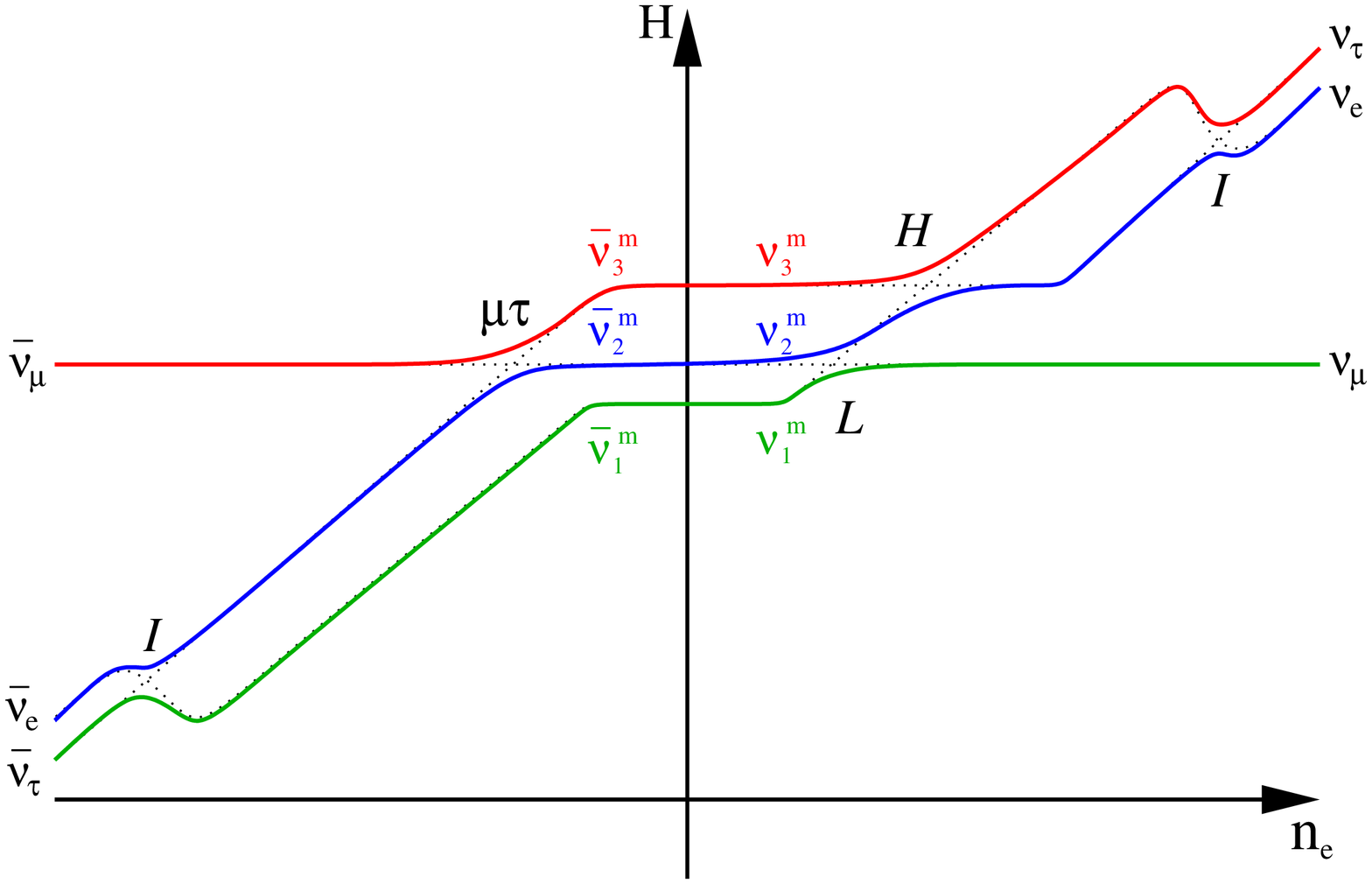}
  \includegraphics[width=0.45\textwidth]{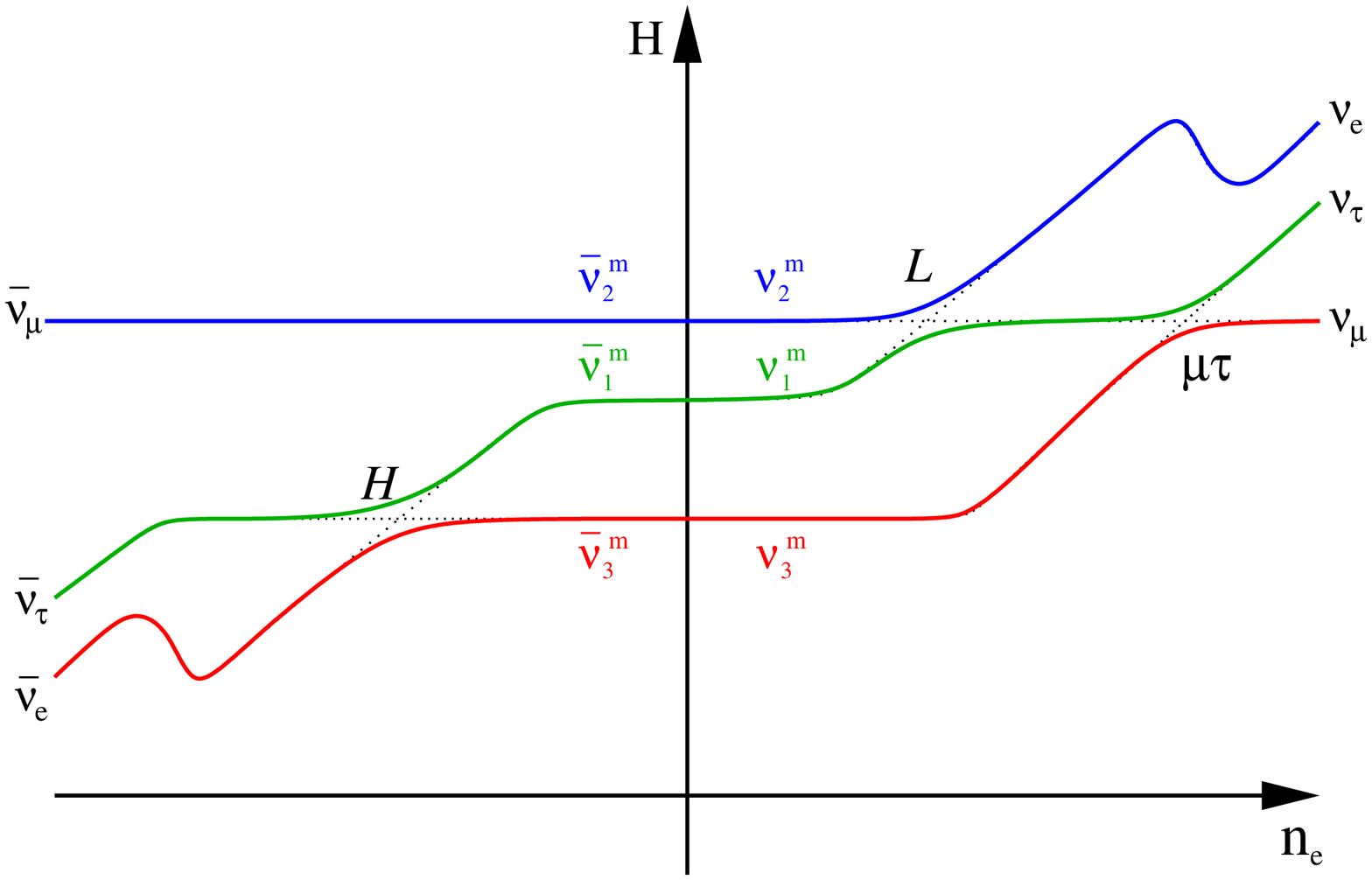}
  \includegraphics[width=0.45\textwidth]{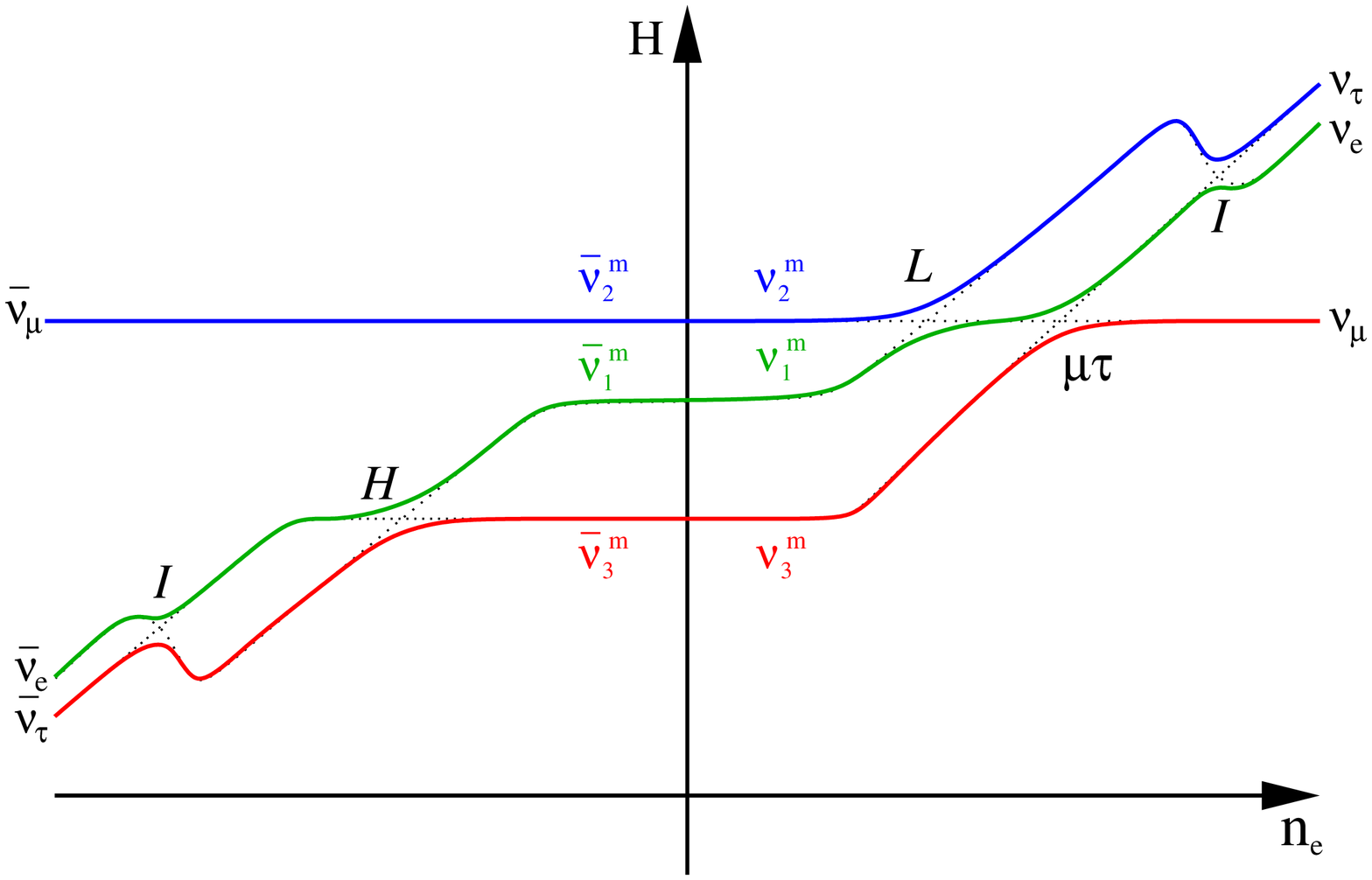}
  \end{center}
  \caption{Level-crossing schemes, first panel is for the case of
    normal hierarchy (oscillations only), the second includes the NSI
    effect.  The two lower panels correspond to the inverse hierarchy,
    oscillations only and oscillations + NSI, respectively.}
  \label{fig:scheme}
\end{figure}

Several comments are in order: First, in contrast to the standard $H$
and $L$-resonances, related to the kinetic term, the density itself
does not explicitly enter into the resonance condition, provided that
the density is high enough to neglect the kinetic terms. Analogously
the energy plays no role in the resonance condition, which is
determined only by the electron fraction $Y_e$. Moreover, in contrast
to the standard resonances, the $I$-resonance occurs for both
neutrinos and antineutrinos simultaneously~\cite{Valle:1987gv}.
Finally, as indicated in Fig.~\ref{fig:scheme}
the $\nu_e$'s ($\bar\nu_e$) are not created as the heaviest (lightest)
state but as the intermediate state, therefore the flavor composition
of the neutrinos arriving at the $H$-resonance is exactly the opposite
of the case without NSI.
As we show in Sec.~\ref{sec:observables}, this fact can lead to
important observational consequences.

In order to calculate the hopping probability between matter
eigenstates at the $I$-resonance we use the Landau-Zener
approximation for two flavors
\begin{equation}
P_{LZ}^{I} \approx e^{-\frac{\pi}{2}\gamma_{I}}~,
%\label{res-i:plz-eq}
\label{eq:plz-i1}
\end{equation}
where $\gamma_{I}$ stands for the adiabaticity parameter, which can be
generally written as
\begin{equation}
\gamma_{I} = \left|\frac{E_2^{\rm m}-E_1^{\rm m}}{2\dot \theta^{\rm
    m}}\right|_{r_{I}}~,
\label{eq:gamma-i1-1}
\end{equation}
where $\dot \theta^{\rm m}\equiv {\rm d}\theta^{\rm m}/{\rm d}r$.
If one applies this formula to the $e-\tau'$ box of Eq.~(\ref{eq:HI1})
assuming that  $\tan 2\theta^{\rm m}_{I} = 2H_{e\tau}'/(H_{\tau\tau}'-H_{ee})$
and $E_2^{\rm m}-E_1^{\rm m} = \left[ (H_{\tau\tau}'-H_{ee})^2 +
4H_{e\tau}' \right]^{1/2}$ 
one gets
\begin{eqnarray}
\gamma_{I} & = & \left| \frac{4H_{e\tau}'^2}{(\dot H_{\tau\tau}'-\dot
    H_{ee})} \right|_{r_{I}} = \left| 
    \frac{16V_0\rho \varepsilon_{e\tau}'^2}{(1+\varepsilon^{I})^3
    \dot Y_e} \right|_{r_{I}} \nonumber \\
            & \approx &
    4\times10^{9} r_{s,5}\rho_{11}
    \varepsilon_{e\tau}'^2 f(\varepsilon^{I})~,
\label{eq:gamma-i1-2}
\end{eqnarray}
where the parametrization of the $Y_e$ profile has been defined as in
Eq.~(\ref{eq:Ye}) with $b=0.16$. The density $\rho_{11}$ represents the
density in units of $10^{11}$~g/cm$^3$, $r_{s,5}$ stands for $r_s$ in
units of $10^5$~cm, and $f(\varepsilon^{I})$ is a function whose value
is of the order $\mathcal{O}(1)$ in the range of parameters we are
interested in.
Taking all these factors into account it follows that the internal
resonance will be adiabatic provided that $\varepsilon_{e\tau}'\gtrsim
10^{-5}$, well below the current limits, in full numerical agreement
with, e.~g., Ref.~\cite{Nunokawa:1996ve}.

\begin{figure}
  \begin{center}
\includegraphics[width=0.3\textwidth,angle=-90]{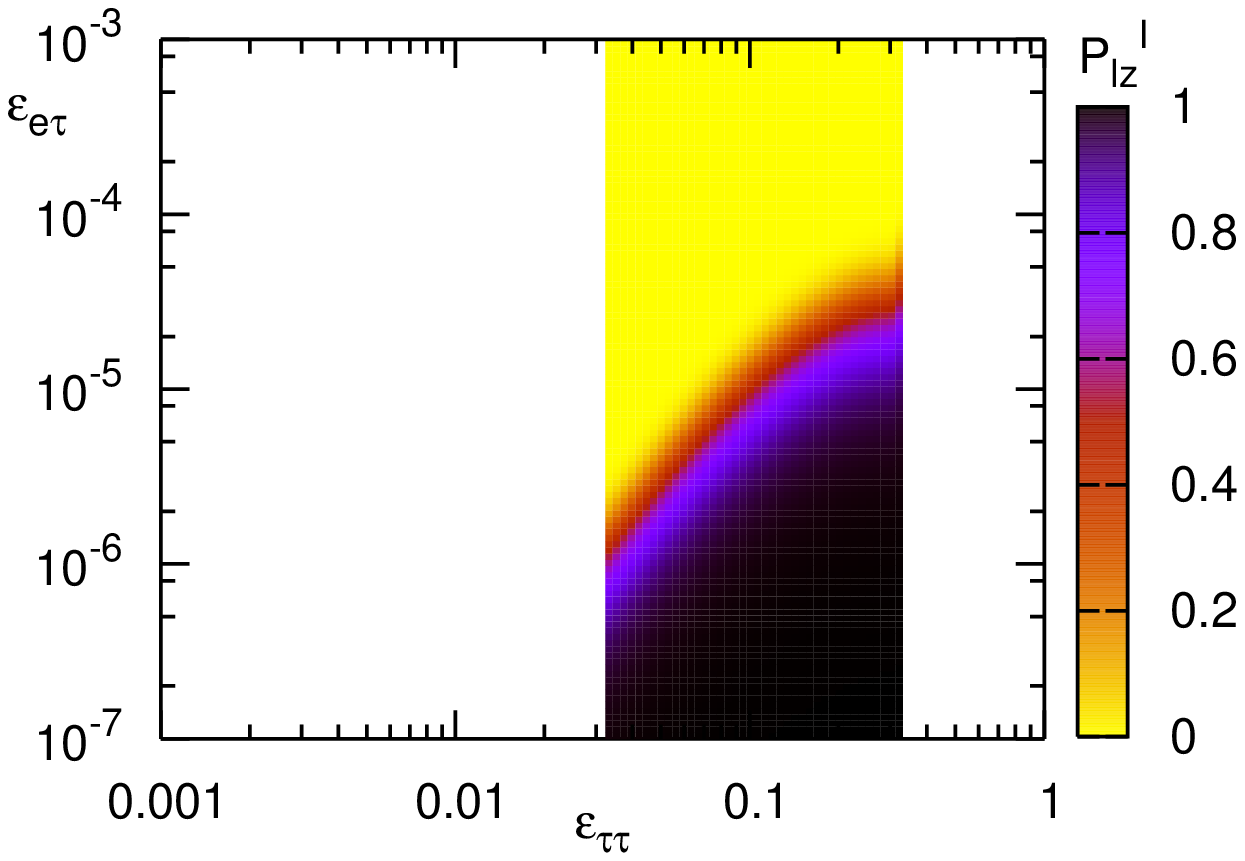}
\includegraphics[width=0.3\textwidth,angle=-90]{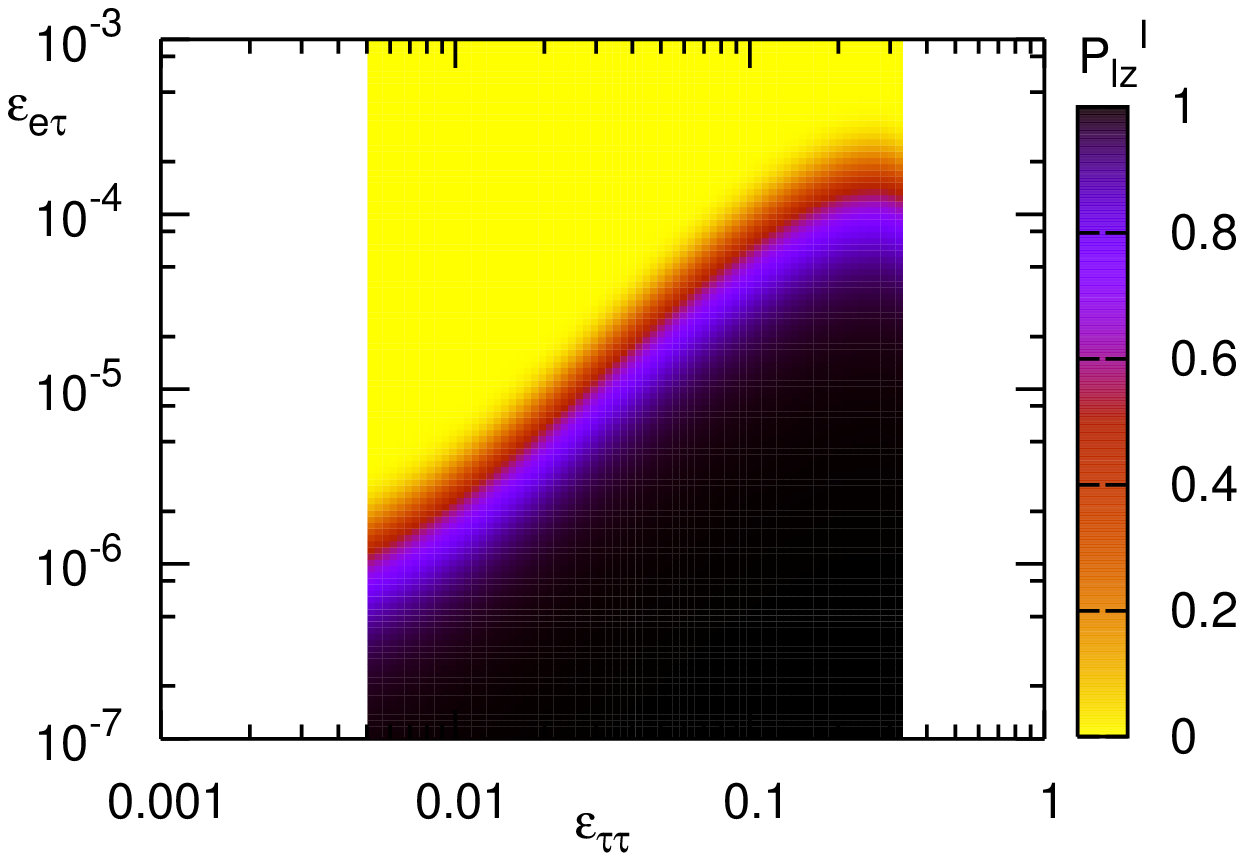}
  \end{center}
  \caption{Contours of constant jump probability at the $I$-resonance
    in terms of $\varepsilon_{\tau\tau}$ and $\varepsilon_{e\tau}$ for
    two profiles corresponding to Fig.~\ref{fig:snprofiles} at $2$~s
    with $a = 0.235$ and $b = 0.175$ (upper panel) and $15.7$~s with
    $a = 0.26$ and $b = 0.195$ (bottom panel). For simplicity the
    other $\varepsilon$'s have been set to zero.}
  \label{fig:adiab_I}
\end{figure}

In Fig.~\ref{fig:adiab_I} we show the resonance condition as well as
the adiabaticity in terms of $\varepsilon_{\tau\tau}$ and
$\varepsilon_{e\tau}$ assuming the other
$\varepsilon_{\alpha\beta}=0$. In order to illustrate the dependence
on time we consider profiles inspired in the numerical profiles of
Fig.~\ref{fig:snprofiles} at $t=2$~s (upper panel) and 15.7 s (bottom
panel). For definiteness we take $Y_e^{\rm min}$ as the electron
fraction at which the density has value of $5\times 10^{11}$g/cm$^3$.
For comparison with Fig.~\ref{fig:YeI-eI} we have assumed $Y_e^{\rm
  min}=10^{-2}$ in the case of 15.7~s. We observe how the border of
adiabaticity depends on $\varepsilon_{\tau\tau}$ through the value of
the density at $r_I$ which in turn depends on time.

%%%%%%%%%%%%%%%%%%%%%%%%%%%%%%%%%%%%%%%%%%%%%%%%%%%%%%%%%%%%%%%%%%%%%%%%

Before moving to the discussion of the outer resonances a comment is
in order, namely, how does the formalism change for other non-standard
interaction models. First note that the whole treatment presented
above also applies to the case of NSI on up-type quarks, except that
the position of the internal resonance shifts with respect to the
down-quark case.  Indeed, in this case the NSI potential
\begin{equation}
(V_{\rm nsi}^{u})_{\alpha\beta} 
=\varepsilon^{u}_{\alpha\beta}V_0\rho(1+Y_e)~,
\end{equation}
would induce a similar internal resonance for the condition
$Y_e=\varepsilon^I/(1-\varepsilon^I)$. 

In contrast, for the case of NSI with electrons, the NSI potential is
proportional to the electron fraction, and therefore no internal
resonance would appear.

%%%%%%%%%%%%%%%%%%%%%%%%%%%%%%%%%%%%%%%%%%%%%%%%%%%%%%%%%%%%%%%%%%%%%%%%

\subsection{Neutrino Evolution in the Outer Regions}
\label{sec:neutr-evol-outer}

%%%%%%%%%%%%%%%%%%%%%%%%%%%%%%%%%%%%%%%%%%%%%%%%%%%%%%%%%%%%%%%%%%%%%%%%

In the outer layers of the SN envelope neutrinos can undergo important
flavor transitions at those points where the matter induced potential
equals the kinetic terms. 
In absence of NSI this condition can be expressed as $V_{CC}\approx
\Delta m^2/(2E)$.  Neutrino oscillation experiments indicate two mass
scales, $\Delta m^2_{\rm atm}$ and $\Delta m^2_\odot\equiv
m_2^2-m_1^2$~\cite{Maltoni:2004ei}, hence two different resonance
layers arise, the so-called $H$-resonance and the $L$-resonance,
respectively.

The presence of NSI with values of $|\varepsilon_{\alpha\beta}|
\lesssim 10^{-2}$ modifies the properties of the $H$ and $L$
transitions~\cite{Mansour:1997fi,Bergmann:1998rg,Fogli:2002xj}. In
particular one finds that the effects of the NSI can be described as
in the standard case by embedding the $\varepsilon$'s into effective
mixing angles~\cite{Fogli:2002xj}. An analogous ``confusion'' between
$\sin\theta_{13}$ and the corresponding NSI parameter
$\varepsilon_{e\tau}$ has been pointed out in the context of
long-baseline neutrino oscillations in
Refs.~\cite{huber:2001de,huber:2002bi}.

In this section we perform a more general and complementary study for
slightly higher values of the NSI parameters:
$|\varepsilon_{\alpha\beta}|\gtrsim {\rm few}~10^{-2}$, still allowed
by current limits, and for which the $I$-resonance could occur.

The phenomenological assumption of hierarchical squared mass
differences, $|\Delta m^2_{\rm atm}|\gg \Delta m^2_\odot$, allows, for
not too large $\varepsilon$'s, a factorization of the 3$\nu$ dynamics
into two 2$\nu$ subsystems roughly decoupled for the $H$ and $L$
transitions~\cite{Kuo:1989qe}.
To isolate the dynamics of the $H$ transition, one usually rotates the 
neutrino flavor basis by $U^\dagger(\theta_{23})$, and extracts the
submatrix with indices (1,3)~\cite{Mansour:1997fi,Fogli:2002xj}.
Whereas this method works perfectly for small values of
$\varepsilon_{\alpha\beta}$ it can be dangerous for values above
$10^{-2}$. In order to analyze how much our case deviates from the
simplest approximation we have performed a rotation with the angle
$\theta_{23}''\equiv \theta_{23}-\alpha$ instead of just
$\theta_{23}$. By requiring that the new rotation diagonalizes the
submatrix (2,3) at the $H$-resonance layer one obtains the following
expression for the correction angle $\alpha$
\begin{eqnarray}
\label{eq:alfa}
\tan(2\alpha) & = & \left[\Delta_{\odot}s2_{12}s_{13} +
  V_{\tau\tau}^{NSI} s2_{23} - 2V_{\mu\tau}^{NSI}
  c2_{23}\right]/\nonumber \\ 
 & & \left[(\Delta_{\rm atm}+ \frac{1}{2}\Delta_{\odot}) 
  c^2_{13}+\frac{1}{4}\Delta_{\odot} c2_{12}(-3+c2_{13}) \right. \nonumber
  \\
& & \left.+ V_{\tau\tau}^{NSI} c2_{23} + 2 V_{\mu\tau}^{NSI}
  s2_{23}\right]~,
\end{eqnarray}
where $\Delta_{\rm atm}\equiv \Delta m^2_{\rm atm}/(2E)$ and
$\Delta_{\odot}\equiv \Delta m^2_\odot/(2E)$. In our notation $s_{ij}$
and $s2_{ij}$ represent $\sin\theta_{ij}$ and $\sin(2\theta_{ij})$,
respectively. The parameters $c_{ij}$ and $c2_{ij}$ are
analogously defined. 
In the absence of NSI $\alpha$ is just a small correction to
$\theta_{23}$~\footnote{Note that, in the limit of high densities one
  recovers the rotation angle obtained for the internal $I$-resonance
  $\theta_{23}'' \to \theta_{23}'$ after neglecting the kinetic terms.
},
\begin{equation}
\tan(2\alpha)\approx
\Delta_{\odot}s2_{12}s_{13}/\Delta_{\rm atm}c^2_{13} \lesssim
\mathcal{O}(10^{-3})~. 
\end{equation}

In order to calculate $\alpha$ we need to know the $H$-resonance
point. To calculate it one can proceed as in the case without NSI,
namely, make the $\theta_{23}''$ rotation and analyze the submatrix
$(1,3)$. The new Hamiltonian $H_{\alpha\beta}''$ has now the form
\begin{eqnarray}
\label{eq:H'_h}
H_{ee}'' & = & V_0\rho [Y_e + \varepsilon_{ee}(2-Y_e)] + \Delta_{\rm
  atm}s^2_{13}  \nonumber \\
        &   & + \Delta_{\odot}(c^2_{13}s^2_{12}+s^2_{13})~, \nonumber\\ 
H_{\tau\tau}'' & = & V_0\rho (2-Y_e)\varepsilon_{\tau\tau}'' + \Delta_{\rm
  atm}c^2_{13}c^2_{\alpha}   \nonumber \\
        &   & + \Delta_{\odot}\left[c^2_{13}c^2_{\alpha} +
  (s_{\alpha}c_{12}+c_{\alpha}s_{12}s_{13})^2 \right]~, \nonumber \\ 
H_{e\tau}'' & = &  V_0\rho (2-Y_e)\varepsilon_{e\tau}'' +
  \frac{1}{2}\Delta_{\rm   atm}s2_{13}c_{\alpha}  \nonumber \\ 
        &   & + \frac{1}{2}\Delta_{\odot}(-c_{13}s_{\alpha}s2_{12}
+c^2_{12}c_{\alpha}s2_{13})~. 
\end{eqnarray} 
We have defined $\varepsilon_{\tau\tau}'' =
\varepsilon_{\tau\tau}c^2_{23-\alpha} +
\varepsilon_{\mu\tau}s2_{23-\alpha}$, and  $\varepsilon_{e\tau}'' =
\varepsilon_{e\tau}c_{23-\alpha} +
\varepsilon_{e\mu}s_{23-\alpha}$, where $s_{23-\alpha}\equiv
\sin(\theta_{23}-\alpha),~c_{23-\alpha}\equiv\cos(\theta_{23}-\alpha)$,
and
$s2_{23-\alpha}\equiv\sin(2\theta_{23}-2\alpha),
~c2_{23-\alpha}\equiv\cos(2\theta_{23}-2\alpha)$.  
The resonance condition for the $H$ transition,
$H_{ee}''=H_{\tau\tau}''$ can be then written as 
\begin{eqnarray}
V_0 \rho^H & [Y_e^H + 
  (\varepsilon_{ee}-\varepsilon_{\tau\tau}'')(2-Y_e^H)] = \Delta_{\rm 
  atm}(c^2_{13} c^2_{\alpha}
-s^2_{13}) & \nonumber \\
& +\Delta_{\odot}[c^2_{12}(c^2_{13}-c^2_{\alpha}s^2_{13})-
  s^2_{\alpha}s^2_{12}+\frac{1}{2}s2_{\alpha}s2_{12} s_{13}] & .
\label{eq:resH}
\end{eqnarray}
It can be easily checked how in the limit of
$\varepsilon_{\alpha\beta} \to 0$ one recovers the standard resonance
condition,
\begin{equation}
V_0 \rho^H Y_e^H\approx
\Delta_{\rm atm}c2_{13}~.
\end{equation}
In the region where the $H$-resonance occurs $Y_e^H\approx 0.5$. 

\begin{figure}
  \begin{center}
    \includegraphics[angle=-0,width=0.45\textwidth]{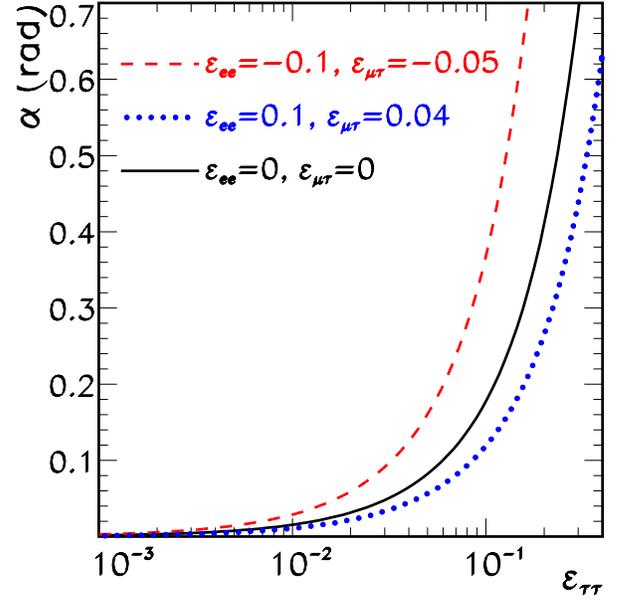}
  \end{center}
  \caption{Angle $\alpha$ as function of $\varepsilon_{\tau\tau}$ for
  different values of $\varepsilon_{ee}$ and $\varepsilon_{\mu\tau}$, 
  in the case of neutrinos of energy $10$~MeV, with normal mass hierarchy, and
  $s^2_{13}=10^{-5}$. The other NSI parameters take the following
  values: $\varepsilon_{e\mu}=0$ and  $\varepsilon_{e\tau}=10^{-3}$.
\label{fig:alpha}}
\end{figure}

Taking into account Eqs.~(\ref{eq:alfa}) and~(\ref{eq:resH}) one can
already estimate how the value of $\alpha$ changes with the NSI
parameters. 
In Fig.~\ref{fig:alpha} we show the dependence of $\alpha$ on the
$\varepsilon_{\tau\tau}$ after fixing the value of the other NSI
parameters. One can see how for $\varepsilon_{\tau\tau}\gtrsim
10^{-2}$ the approximation of neglecting $\alpha$ significantly
worsens.  Assuming $\theta_{23}=\pi/4$ and a fixed value of
$\varepsilon_{\mu\tau}$ one can easily see that
$\varepsilon_{\tau\tau}$ basically affects the numerator in
Eq.~(\ref{eq:alfa}). Therefore one expects a rise of $\alpha$ as the
value of $\varepsilon_{\tau\tau}$ increases, as seen in
Fig.~\ref{fig:alpha}.  The dependence of $\alpha$ on
$\varepsilon_{\mu\tau}$ is correlated to the relative sign of the mass
hierarchy and $\varepsilon_{\mu\tau}$. For instance, for normal mass
hierarchy and positive values of $\varepsilon_{\mu\tau}$ the
dependence is inverse, namely, higher values of
$\varepsilon_{\mu\tau}$ lead to a suppression of $\alpha$.  Apart from
this general behavior, $\alpha$ also depends on the diagonal term
$\varepsilon_{ee}$ as seen in Fig.~\ref{fig:alpha}. This effect occurs
by shifting the resonance point through the resonance condition in
Eq.~(\ref{eq:resH}).

One can now calculate the jump probability between matter eigenstates
in analogy to the $I$-resonance by means of the Landau-Zener
approximation, see Eqs.~(\ref{eq:plz-i1}),~(\ref{eq:gamma-i1-1}),
and~\ref{eq:gamma-i1-2},
\begin{equation}
P_{LZ}^{H} \approx e^{-\frac{\pi}{2}\gamma_{H}}~,
\label{eq:plz-h}
\end{equation}
where $\gamma_{H}$ represents the adiabaticity parameter at the
$H$-resonance, which can be written as
\begin{equation}
\gamma_{H} =  \left| \frac{4H_{e\tau}''^2}{(\dot H_{\tau\tau}''-\dot
    H_{ee}'')} \right|_{r_{H}}~,
\label{eq:gammaH}
\end{equation}
where the expressions for $H_{\alpha\beta}''$ are given in
Eqs~(\ref{eq:H'_h}).  

\begin{figure}
  \begin{center}
  \includegraphics[angle=-0,width=0.45\textwidth]{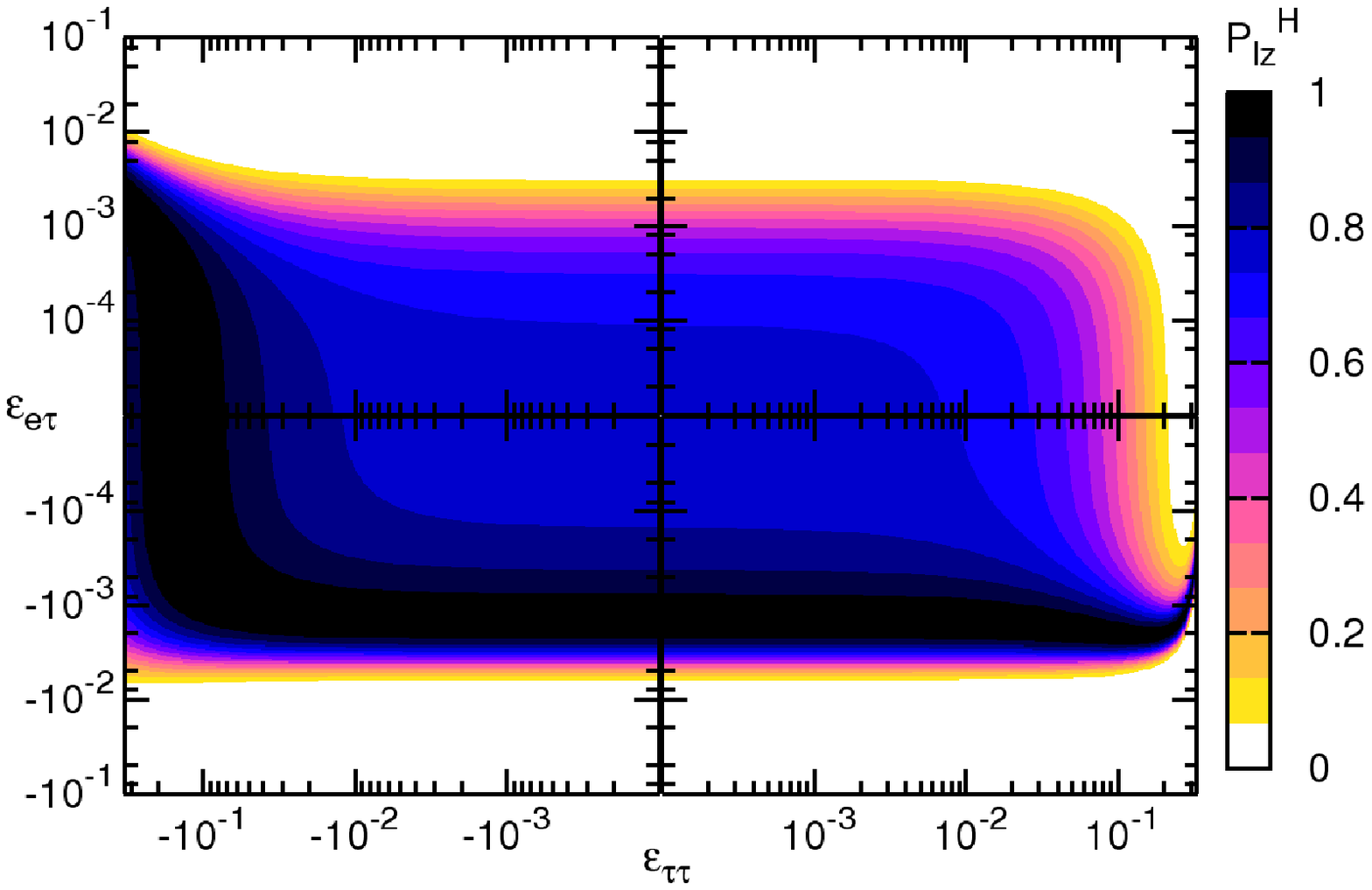} %%
   \includegraphics[angle=-0,width=0.45\textwidth]{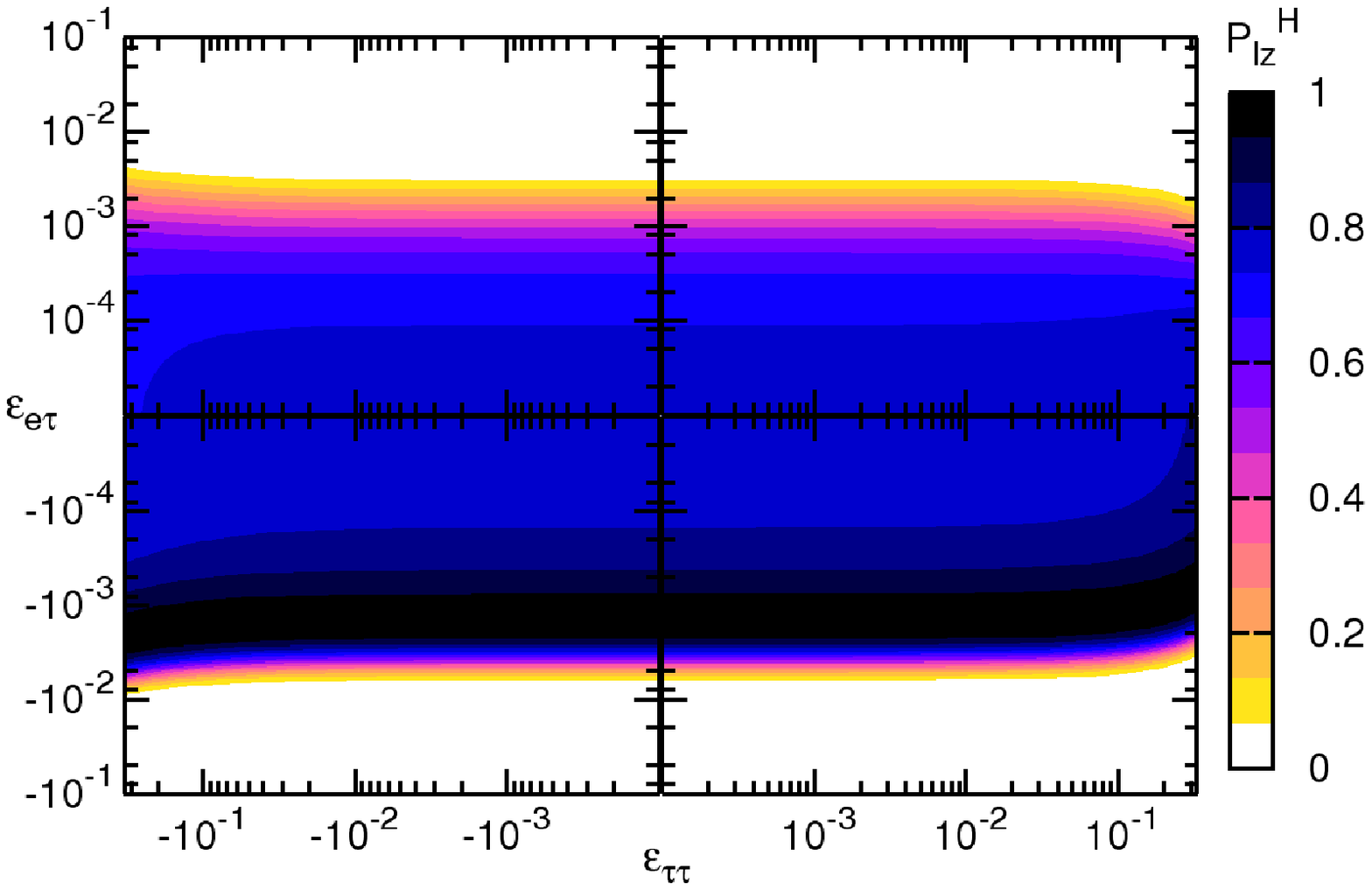} %%
  \end{center}
  \caption{Landau-Zener jump probability isocontours at the
    $H$-resonance in terms of $\varepsilon_{e\tau}$ and
    $\varepsilon_{\tau\tau}$ for $10$~MeV antineutrinos in the case of
    inverted mass hierarchy. Upper panel: $\alpha$ given by
    Eq.~(\ref{eq:alfa}). Bottom panel: $\alpha$ set to zero. 
  The remaining parameters take the following 
values: $\sin^2\theta_{13}=10^{-5},~\varepsilon_{e\tau}=10^{-3},
~\varepsilon_{ee}=\varepsilon_{e\mu}=0$. See text.
  }
\label{fig:p-h}
\end{figure}

Let us first consider the case $|\varepsilon_{\alpha\beta}|\lesssim
10^{-2}$.  In this case $\alpha\approx 0$ and one can rewrite the
adiabaticity parameter as
\begin{equation}
\gamma_{H} \approx \frac{\Delta_{\rm
    atm}\sin^2(2\theta_{13}^{eff})}{\cos(2\theta_{13}^{eff}) |{\rm
    d}\ln V /{\rm d}r|_{r_H}}~, 
\label{eq:gammaH_lisi}
\end{equation}
where 
\begin{equation}
  \label{eq:deg}
  \theta_{13}^{eff} = \theta_{13} + \varepsilon_{e\tau}''
  (2-Y_e)/Y_e  
\end{equation}
in agreement with Ref.~\cite{Fogli:2002xj}.
For slightly larger $\varepsilon$'s there can be significant
differences. In Fig.~\ref{fig:p-h} we show $P_{LZ}^{H}$ in the
$\varepsilon_{e\tau}$-$\varepsilon_{\tau\tau}$ plane for antineutrinos
with energy $10$~MeV in the case of inverse mass hierarchy, using
Eq.~(\ref{eq:plz-h}) with (upper panel) and without (bottom panel) the
$\alpha$ correction.  The values of $\theta_{13}$ and
$\varepsilon_{e\tau}$ have been chosen so that the jump probability
lies in the transition regime between adiabatic and strongly non
adiabatic.
In the limit of small $\varepsilon_{\tau\tau}$, $\alpha$ becomes
negligible and therefore both results coincide. From
Eq.~(\ref{eq:gammaH_lisi}) one sees how as the value of
$\varepsilon_{e\tau}$ increases $\gamma_H$ gets larger and therefore
the transition becomes more and more adiabatic. For negative values of
$\varepsilon_{e\tau}$ there can be a cancellation between
$\varepsilon_{e\tau}$ and $\theta_{13}$, and as a result the
transition becomes non-adiabatic.

\begin{figure}
  \begin{center}
\includegraphics[width=0.48\textwidth,height=7.cm,angle=0]{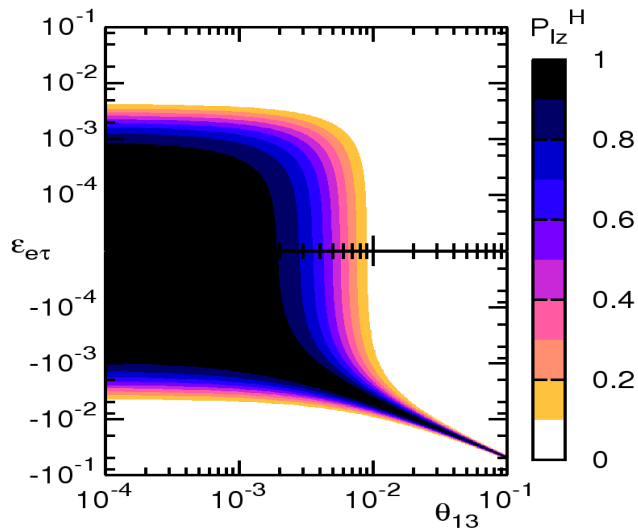}
  \end{center}
\caption{Landau-Zener jump probability isocontours at the
  $H$-resonance in terms of $\varepsilon_{e\tau}$ and $\theta_{13}$
  for $\varepsilon_{\tau\tau}=10^{-4}$. Antineutrinos with energy
  $10$~MeV and inverted mass hierarchy has been assumed.}
  \label{fig:confusion}
\end{figure}

An additional consequence of Eq.~(\ref{eq:deg}) is that a degeneracy
between $\varepsilon_{e\tau}$ and $\theta_{13}$ arises. This is seen
in Fig.~\ref{fig:confusion}, which gives the contours of $P_{\rm
  LZ}^H$ in terms of $\varepsilon_{e\tau}$ and $\theta_{13}$ for
$\varepsilon_{\tau\tau}=10^{-4}$. One sees clearly that the same
Landau-Zener hopping probability is obtained for different
combinations of $\varepsilon_{e\tau}$ and $\theta_{13}$. This leads to
an intrinsic ``confusion'' between the mixing angle and the
corresponding NSI parameter, which can not be disentangled only in the
context of SN neutrinos, as noted in Ref.~\cite{Fogli:2002xj}.
 
We now turn to the case of $|\varepsilon_{\tau\tau}| \geq 10^{-2}$.  As
$|\varepsilon_{\tau\tau}|$ increases the role of $\alpha$ becomes
relevant.  Whereas in the bottom panel $P_{LZ}^{H}$ remains basically
independent of $\varepsilon_{\tau\tau}$, one can see how in the upper
panel $P_{LZ}^{H}$ becomes strongly sensitive to
$\varepsilon_{\tau\tau}$ for $|\varepsilon_{\tau\tau}| \geq 10^{-2}$.

One sees that for positive values of $\varepsilon_{\tau\tau}$ it tends
to adiabaticity whereas for negative values to non-adiabaticity. This
follows from the dependence of $H_{e\tau}''$ on $\alpha$, essentially
through the term $-\Delta_\odot c_{13}s_\alpha s2_{12}$, see
Eq.~(\ref{eq:H'_h}).  For $|\varepsilon_{\tau\tau}| \geq 10^{-2}$ one sees
that $\sin\alpha$ starts being important, and as a result this term
eventually becomes of the same order as the others in $H_{e\tau}''$.
At this point the sign of $\varepsilon_{\tau\tau}$, and so the sign of
$\sin\alpha$, is crucial since it may contribute to the enhancement or
reduction of $H_{e\tau}''$. This directly translates into a trend
towards adiabaticity or non-adiabaticity, seen in Fig.~\ref{fig:p-h}.
Thus, for the range of $\varepsilon_{\tau\tau}$ relevant for the
NSI-induced internal resonance the adiabaticity of the outer $H$
resonance can be affected in a non-trivial way. 

Turning to the case of the $L$ transition a similar expression can be
obtained by rotating the original Hamiltonian by
$U(\theta_{13})^\dagger
U(\theta_{23})^\dagger$~\cite{Mansour:1997fi,Fogli:2002xj}. However,
in contrast to the case of the $H$-resonance, where the mixing angle
$\theta_{13}$ is still unknown, in the case of the $L$ transition the
angle $\theta_{12}$ has been shown by solar and reactor neutrino
experiments to be large~\cite{Maltoni:2004ei}. As a result, for the
mass scale $\Delta_\odot$ this transition will always be adiabatic
irrespective of the values of $\varepsilon_{\alpha\beta}$, and will
affect only neutrinos.

%%%%%%%%%%%%%%%%%%%%%%%%%%%%%%%%%%%%%%%%%%%%%%%%%%%%%%%%%%%%%%%%%%%%%%
\section{Observables and sensitivity}
\label{sec:observables}
%%%%%%%%%%%%%%%%%%%%%%%%%%%%%%%%%%%%%%%%%%%%%%%%%%%%%%%%%%%%%%%%%%%%%%

As mentioned in the introduction one of the major motivations to study
NSI using the neutrinos emitted in a SN is the enhancement
of the NSI effects on the neutrino propagation through the SN envelope
due to the specific extreme matter conditions that characterize it.
In this section we analyze how these effects translate into observable
effects in the case of a future galactic SN.

 Schematically, the neutrino emission by a SN can be divided into four
 stages: Infall phase, neutronization burst, accretion, and
 Kelvin-Helmholtz cooling phase.
 During the infall phase and neutronization burst only $\nu_e$'s are
 emitted, while the bulk of neutrino emission is released in all
 flavors in the last two phases.
 Whereas the neutrino emission characteristics of the two initial
 stages are basically independent of the features of the progenitor,
 such as the core mass or equation of state (EoS), the details of the
 neutrino spectra and luminosity during the accretion and cooling
 phases may significantly change for different progenitor models.
 As a result, a straightforward extraction of oscillation parameters
 from the bulk of the SN neutrino signal seems hopeless. Only features
 in the detected neutrino spectra which are independent of unknown SN
 parameters should be used in such an
 analysis~\cite{Kachelriess:2004vs}.

%% vvv

 The question then arises as to how can one obtain information about
 the NSI parameters.
 Taking into account that the main effect of NSI is to generate new
 internal neutrino flavor transitions, one possibility is to invoke
 theoretical arguments that involve different aspects of the SN
 internal dynamics.

 In Ref.~\cite{Nunokawa:1996ve} it was argued that such an internal
 flavor conversion during the first second after the core bounce might
 play a positive role in the so-called SN shock reheating problem. It
 is observed in numerical
 simulations~\cite{Liebendoerfer:2000cq,Rampp:2002bq,Thompson:2002mw,Sumiyoshi:2005ri}
 that as the shock wave propagates it loses energy until it gets
 stalled at a few hundred km. It is currently believed that after
 neutrinos escape the SN core they can to some extent deposit energy
 right behind and help the shock wave continue outwards. On the other
 hand it is also believed that due to the composition in matter of the
 protoneutronstar (PNS) the mean energies of the different neutrino
 spectra obey $\langle E_{\nu_e} \rangle < \langle E_{\bar\nu_e}
 \rangle < \langle E_{\nu_\mu,\nu_\tau} \rangle$. This means that a
 resonant conversion between $\nu_e (\bar\nu_e)$ and $\nu_{\mu,\tau}
 (\bar\nu_{\mu,\tau})$ between the neutrinosphere and the position of
 the stalled shock wave would make the $\nu_e (\bar\nu_e)$ spectra
 harder, and therefore the energy deposition would be larger, giving
 rise to a shock wave regeneration effect.

 Another argument used in the literature was the possibility that the
 $r-$process nucleosynthesis, responsible for synthesizing about half
 of the heavy elements with mass number $A>70$ in nature, could occur
 in the region above the neutrinosphere in
 SNe~~\cite{Qian:2003wd,Pruet:2004vb}. A necessary condition is
 $Y_e<0.5$ in the nucleosynthesis region. The value of the electron
 fraction depends on the neutrino absorption rates, which are
 determined in turn by the $\nu_e(\bar\nu_e)$ luminosities and energy
 distribution. These can be altered by flavor conversion in the inner
 layers due to the presence of NSI. Therefore by requiring the
 electron fraction be below 0.5 one can get information about the
 values of the NSI parameters.

 While it is commonly accepted that neutrinos will play a crucial role
 in both  the shock wave re-heating as well as the $r-$process
 nucleosynthesis, there are still other astrophysical factors
 that can affect both.
 While the issue remains under debate we prefer to stick to arguments
 directly related to physical observables in a large water Cherenkov
 detector. There are several possibilities.

 \begin{itemize}

 \item[(A)] the modulations in the $\bar\nu_e$ spectra due to the
   passage of shock waves through the
   supernova~\cite{Schirato:2002tg,Fogli:2003dw,Tomas:2004gr}
 \item[(B)] the modulation in the $\bar\nu_e$ spectra due to the time
   dependence of the electron fraction, induced by the $I$-resonance
 \item[(C)]  the modulations in the $\bar\nu_e$ spectra due to the Earth
   matter~\cite{Lunardini:2001pb,Dighe:2003be,Dighe:2003jg,Dighe:2003vm}
\item[(D)]  detectability of the neutronization $\nu_e$
   burst~\cite{Takahashi:2003rn,Kachelriess:2004ds}
 \end{itemize}
 Three of these observables, 1, 3 and 4 have already been considered
 in the literature in the context of neutrino oscillations. 
 Here we discuss the potential of the above promising observables in
 providing information about the NSI parameters. It is important to
 pay attention to the possible ocurrence of the internal $I$-resonance
 and to its effect in the external $H$ and $L$-resonances. The first
 can induce a genuinely new observable effect, item 2 above. 

 Here we concentrate on neutral current-type non-standard
 interactions, hence there will be not effect in the main reaction in
 water Cherenkov and scintillator detectors, namely the inverse beta
 decay, $\bar\nu_e+p\to e^++n$~\footnote{For the case of NSI with
   electrons both the vector and axial components of
   $\varepsilon_{\alpha\beta}^e$ will contribute to the $\nu-e$ cross
   section.}.  For definiteness we take NSI with $d$ (down) quarks, in
 which case the NSI effects will be confined to the neutrino evolution
 inside the SN and the Earth, through the vector component of the
 interaction.

 From all possible combinations of NSI parameters we will concentrate
 on those for which the internal $I$ transition does take place,
 namely $|\varepsilon^{I}|\gtrsim 10^{-2}$, see Fig.~\ref{fig:YeI-eI}.
 Concerning the FC NSI parameters we will consider
 $|\varepsilon_{e\tau}'|$ between ${\rm few}\times 10^{-5}$ and
 $10^{-2}$, range in which the $I$-resonance is adiabatic, see
 Fig.~\ref{fig:adiab_I}.
 In the following discussion we will focus on the extreme cases
 defined in Table~\ref{table:nuschemes}. One of the motivations for
 considering these cases is the fact that the resonances involved
 become either adiabatic or strongly non adiabatic, and hence the
 survival probabilities in the absence of Earth effects or shock wave
 passage, become energy independent. This assumption simplifies the
 task of relating the observables with the neutrino schemes.
\begin{table}
\begin{center}
\begin{tabular}{|c|c|c|c|c|c|}
\hline
 Scheme & Hierarchy & $\sin^2\theta_{13}$ & NSI & $P_{\rm surv}$ & $\bar P_{\rm surv}$  \\
\hline
\hline
$A$ & normal & $\gtrsim 10^{-4}$ & No & 0 & $\cos^2\theta_{12}$ \\
\hline
$B$ & inverted & $\gtrsim 10^{-4}$ & No & $\sin^2\theta_{12}$ & 0 \\ 
\hline
$C$ & any & $\lesssim 10^{-6}$ & No & $\sin^2\theta_{12}$ &
 $\cos^2\theta_{12}$\\
\hline
\hline
$AI$ & normal & $\gtrsim 10^{-4}$ & Yes & $\sin^2\theta_{12}$ &
 $\sin^2\theta_{12}$ \\
\hline
$BI$ & inverted & $\gtrsim 10^{-4}$ & Yes & $\cos^2\theta_{12}$ & $\cos^2\theta_{12}$ \\ 
\hline
$CIa$ & normal & $\lesssim 10^{-6}$ & Yes & 0 & $\sin^2\theta_{12}$ \\
\hline
$CIb$ & inverted & $\lesssim 10^{-6}$ & Yes & $\cos^2\theta_{12}$ & 0  \\
\hline
\end{tabular}
\end{center}
\caption{Definition of the neutrino schemes considered in terms of the
  hierarchy, the value of $\theta_{13}$, and the presence of NSI, as
  described in the text. The values of the survival probabilities for
  $\nu_e$ ($P_{\rm surv}$) and $\bar\nu_e$ ($\bar P_{\rm surv}$) for
  each case are also indicated.}
\label{table:nuschemes}
\end{table}

%%%%%%%%%%%%%%%%%%%%%%%%%%%%%%%%%%%%%%%%%%%%%%%%%%%%%%%%%%%%%%%%%%%%%%
\subsection{Shock wave propagation}
\label{subsec:shockwave}
%%%%%%%%%%%%%%%%%%%%%%%%%%%%%%%%%%%%%%%%%%%%%%%%%%%%%%%%%%%%%%%%%%%%%%

\begin{figure}
  \begin{center}
    \includegraphics[width=0.45\textwidth]{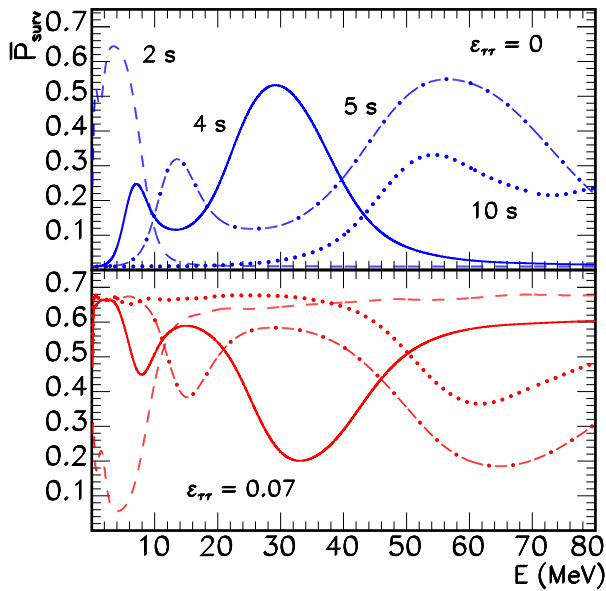}
  \end{center}
  \caption{Survival probability $\bar P_{\rm surv}(E,t)$ for
    $\bar\nu_e$ as function of energy at different times averaged in
    energies with the energy resolution of Super-Kamiokande; for the
    profile shown in Fig.~\ref{fig:snprofiles}. Upper panel: case $B$ is
    assumed for $\sin^2\theta_{13}= 10^{-2}$. Bottom panel: case $BI$,
    with $\varepsilon_{\tau\tau}=0.07,~\varepsilon_{e\tau}= 10^{-4} $
    and the rest of NSI parameters put to zero.}
  \label{fig:psurv}
\end{figure}

During approximately the first two seconds after the core bounce, the
neutrino survival probabilities are constant in time and in energy for
all cases mentioned in Table~\ref{table:nuschemes}. Only the Earth
effects could introduce an energy dependence.
However, at $t\approx 2$ s the $H$-resonance layer is reached by the
outgoing shock wave, see Fig.~\ref{fig:snprofiles}.  The way the shock
wave passage affects the neutrino propagation strongly depends on the
neutrino mixing scenario. In the absence of NSI cases $A$ and $C$ will
not show any evidence of shock wave propagation in the observed
$\bar\nu_e$ spectrum, either because there is no resonance in the
antineutrino channel as in scenario $A$, or because the $H$-resonance
is always strongly non-adiabatic as in scenario $C$.  However, in
scenario $B$, the sudden change in density breaks the adiabaticity of
the resonance, leading to a time and energy dependence of the electron
antineutrino survival probability $\bar P_{\rm surv}(E,t)$. In the
upper panel of Fig.~\ref{fig:psurv} we show $\bar P_{\rm surv}(E,t)$
in the particular case that two shock waves are present, one forward
and a reverse one~\cite{Tomas:2004gr}.  The presence of the shocks
results in the appearance of bumps in survival probability at those
energies for which the resonance region is passed by the shock waves.
All these structures move in time towards higher energies, as the
shock waves reach regions with lower density, leading to observable
consequences in the $\bar\nu_e$ spectrum.

We now turn to the case where NSI are present, which opens the
possibility of internal resonances.  When such $I$-resonance is
adiabatic the situation will be similar to the case without NSI.  For
normal mass hierarchy, $AI$ and $CIa$, $\bar\nu_e$ will not feel the
$H$-resonance and therefore the adiabaticity-breaking effect will not
basically alter their propagation. In contrast, for inverted mass
hierarchy and large $\theta_{13}$, case $BI$, the $H$-resonance occurs
in the antineutrino channel and therefore $\bar\nu_e$ will feel the
shock wave passage.
However, in contrast to case $B$ now $\bar\nu_e$ will reach the
$H$-resonance in a different matter eigenstate: $\bar\nu_1^m$ instead
of $\bar\nu_3^m$, see Fig.~\ref{fig:scheme}. That means that before
the shock wave reaches the $H$-resonance the $\bar\nu_e$ survival
probability will be $\bar P_{\rm surv}\approx \cos^2\theta_{12}\approx
0.7$. Once the adiabaticity of the $H$-resonance is broken by the shock
wave then $\bar\nu_e$ will partly leave as $\bar\nu_3^m$ and therefore
the survival probability will decrease.
As a consequence one expects a pattern in time and energy for the
survival probability in the case $BI$ to be roughly {\em opposite}
than in the case $B$, see bottom panel of Fig.~\ref{fig:psurv}. The
position of the peaks and dips en each panel do not exactly coincide
as the value of $\varepsilon_{\tau\tau}$ roughly shifts the position
of the $H$-resonance.

\begin{figure}
  \begin{center}
\includegraphics[width=0.5\textwidth]{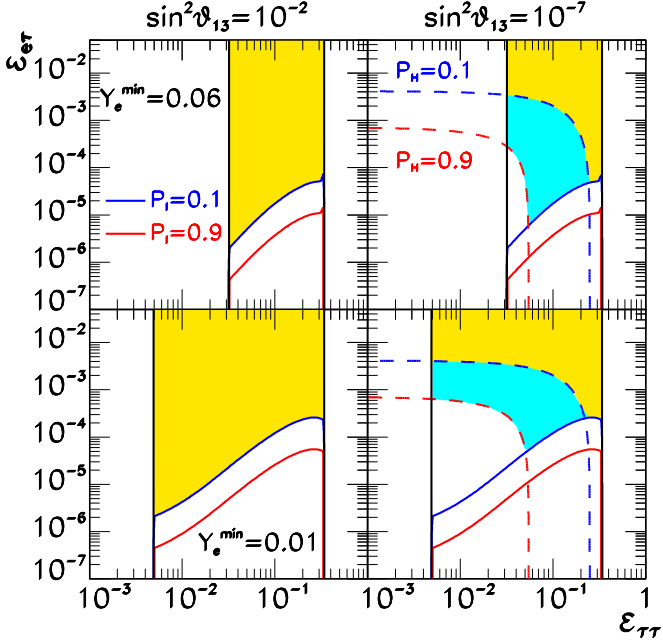}
  \end{center}
  \caption{Range of $\varepsilon_{\tau\tau}$ and $\varepsilon_{e\tau}$
    for which the effect of the shock wave will be observed. In the
    upper panels a minimum value of $Y_e^{\rm min}=0.06$ based on the
    numerical profiles at $t=2$~s has been assumed, see
    Fig.~\ref{fig:snprofiles}. In the lower panels we have considered
    a case with $Y_e^{\rm min}=0.01$ inspired in the profile at
    $t=15.7$~s. The value of $\sin^2\theta_{13}$ has been assumed to
    be $10^{-2}$ and $10^{-7}$ in the left and right panels,
    respectively.  We have also superimposed isocontours of constant
    hopping probability $0.1$ (blue) and $0.9$ (red) in the $I$ (solid
    lines) and $H$ (dashed lines) resonances for inverted mass
    hierarchy and $E=10$~MeV and antineutrinos. The area in yellow
    represents the parameter space where both resonances will be
    adiabatic. In the cyan area the $I$-resonance is assumed to be
    adiabatic whereas $H$ lies in the transition region.}
  \label{fig:res_IH}
\end{figure}

In the left panels of Fig.~\ref{fig:res_IH} we represent in
light-shaded (yellow) the range of $\varepsilon_{e\tau}$ and
$\varepsilon_{\tau\tau}$ for which this {\em opposite} shock wave
imprint would be observable. In the upper panels we have assumed a
minimum value of the electron fraction of $0.06$, based on the
numerical profiles at $t=2$~s of Fig.~\ref{fig:snprofiles}. In the
bottom panels $Y_e^{\rm min}$ is set to $0.01$, inspired in the
profiles at $t=15.7$~s. It can be seen how as time goes on the range
of $\varepsilon_{\tau\tau}$'s for which the $I$-resonance takes place
widens towards to smaller and smaller values.  This is a direct
consequence of the steady deleptonization of the inner layers.

For smaller $\theta_{13}$, case $CIb$, the situation is different.
Except for relatively large $\varepsilon_{e\tau}$ values the
$H$-resonance will be strongly non-adiabatic, as in case $C$.
Therefore the passage of the shock waves will not significantly change
the $\bar\nu_e$ survival probability and will not lead to any
observable effect.
In the right panels of Fig.~\ref{fig:res_IH} we show the same as in
the left panels but for $\sin^2\theta_{13}=10^{-7}$. Whereas for large
values of $\theta_{13}$, left panels, the $H$-resonance is always
adiabatic and one has only to ensure the adiabaticity of the
$I$-resonance, for smaller values of $\theta_{13}$ the adiabaticity of
the $H$-resonance strongly depends on the values of
$\varepsilon_{e\tau}$ and $\varepsilon_{\tau\tau}$, as discussed in
Sec.~\ref{sec:neutr-evol-outer}. This can be seen as a significant
reduction of the yellow area. Only large values of either
$\varepsilon_{e\tau}$ or $\varepsilon_{\tau\tau}$ would still allow
for a clear identification of the {\em opposite} shock wave effects.
In dark-shaded (cyan) we show the region of parameters for which $P_H$
lies in the transition region between adiabatic and strongly
non-adiabatic, and therefore could still lead to some effect.

\begin{figure}
  \begin{center}
\includegraphics[width=0.5\textwidth]{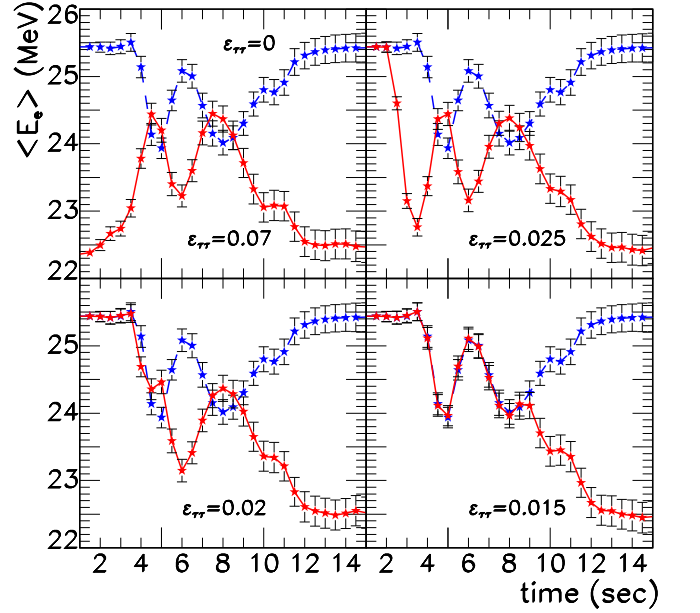}
  \end{center}
  \caption{The average energy of $\bar\nu p\to ne^+$ events binned in
    time for case $B$ (dashed blue) and $BI$ (solid red). In each
    panel different values of $\varepsilon_{\tau\tau}$ have been
    assumed.  The error bars represent 1~$\sigma$ errors in any bin.
    $\varepsilon_{e\tau}= 10^{-4}$.}
  \label{fig:shockwave}
\end{figure}

A useful observable to detect effects of the shock propagation is the
average of the measured positron energies, $\langle E_e\rangle$,
produced in inverse beta decays.  In Fig.~\ref{fig:shockwave}, we show
$\langle E_e \rangle$ together with the one sigma errors expected for
a Megaton water Cherenkov detector and a SN at 10~kpc distance, with a
time binning of 0.5~s, for different neutrino schemes: case $B$ and
case $BI$ with different values of $\varepsilon_{\tau\tau}$. For the
neutrino fluxes we assumed the parametrization given by
Refs.~\cite{MKeil,Keil:2002in} with $\langle E_0(\bar\nu_e) \rangle =
15$~MeV and $\langle E_0(\bar\nu_{\mu,\tau}) \rangle = 18$~MeV and the
following ratio of the total neutrino fluxes
$\Phi_0(\bar\nu_e)/\Phi_0(\bar\nu_{\mu,\tau})=0.8$~\footnote{We assume
that for the values of the NSI parameters considered the initial
neutrino spectra do not significantly change.}.

One can see how the features of the average positron energy are a
direct consequence of the shape of the survival probability, where
dips have to be translated into bumps and vice-versa.

Thus, it is important to stress that whereas in case $B$ one expects the
presence of one or two dips (depending on the structure of the shock
wave, see Ref~\cite{Tomas:2004gr}), or nothing in the other cases, one
or two bumps are expected in case $BI$, as seen in the upper left panel of
Fig.~\ref{fig:shockwave}.  As discussed in Ref.~\cite{Tomas:2004gr}
the details of the dips/bump will depend on the exact shape of the
neutrino fluxes, but as long as general reasonable assumptions like
$\langle E_{\bar\nu_e} \rangle \lesssim \langle E_{\bar\nu_{\mu,\tau}}
\rangle$ are considered the dips/bumps should be observed.

%%%%%%%%%%%%%%%%%%%%%%%%%%%%%%%%%%%%%%%%%%%%%%%%%%%%%%%%%%%%%%%%%%%%%%

\subsection{Time variation of $Y_e$}
\label{subsec:time-variation-y_e}

%%%%%%%%%%%%%%%%%%%%%%%%%%%%%%%%%%%%%%%%%%%%%%%%%%%%%%%%%%%%%%%%%%%%%%

We have just seen how the distorsion of the density profile due to the
shock wave passage through the outer SN envelope can induce a
time-dependent modulation in the $\bar\nu_e$ spectrum in cases $B$ and
$BI$.  However the time dependence of the electron fraction $Y_e$ can
also reveal the presence of NSI leaving a clear imprint in the
observed $\bar\nu_e$ spectrum, as we now explain.

As discussed in Sec.~\ref{subsec:nu_evolution_inner} the region of NSI
parameters leading to $I$-resonance is basically determined by the
minimum and maximum values of the electron fraction, $Y_e^{\rm min}$
and $Y_e^{\rm max}$.  The crucial point is that as the deleptonization
of the proto-neutron star goes on, the value of $Y_e^{\rm min}$
steadily decreases with time. As a result, the range of NSI strengths
for which the $I$-resonance takes place increases with time, as can be
seen in Fig.~\ref{fig:YeI-eI}.

Let us first discuss the observational consequences of the time
dependence of the electron fraction in case $BI$. 
If $\varepsilon_{\tau\tau}$ ($\varepsilon^I$ in general) is large
enough the $I$-resonance will take place right after the core bounce.
In this case, as seen in the upper left panel of
Fig.~\ref{fig:shockwave} the two bumps we have just discussed in
Sec.~\ref{subsec:shockwave} would be clearly observed.  However for
smaller NSI parameter values it could happen that the $I$-resonance
occurs only after several seconds. In particular for the specific
$Y_e$ profile considered we show how this delay could be of roughly 2,
4 or 9 sec for values of $\varepsilon_{\tau\tau}$ of 0.025, 0.02 or
0.015, respectively, see last three panels Fig.~\ref{fig:shockwave}.
As can be inferred from the figure this delay effect can lead to
misidentification of the pure NSI effect. So, for instance, in the
upper right panel, one sees how the two bumps might also be
interpreted as two dips, given the astrophysical uncertainties.  
This subtle degeneracy can only be solved by extra information on, for
example, the time dependence of the spectra or the velocity of the
shock wave. Given the supernova model, however, the time structure of
the signal could eventually not only point out the presence of NSI but
even potentially indicate a range of NSI parameters.

\begin{figure}
  \begin{center}
    \includegraphics[width=0.45\textwidth]{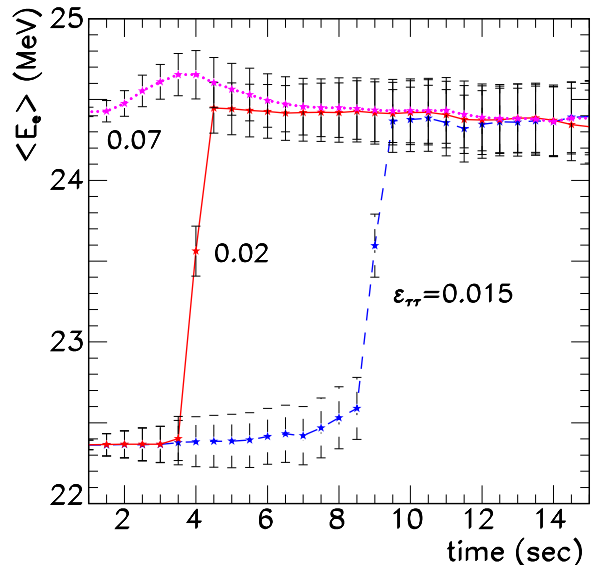}
  \end{center}
  \caption{The average energy of $\bar\nu p\to ne^+$ events binned in
    time for case $AI$ and $CIa$ and different values of
    $\varepsilon_{\tau\tau}$. 
  The error bars represent 1~$\sigma$ errors in any bin.
    $\varepsilon_{e\tau}= 10^{-4}$.
}
  \label{fig:new2}
\end{figure}

Let us now turn to  the normal mass hierarchy scenario (cases $AI$
and $CIa$). In analogy to the $BI$ case, if $\varepsilon^I$ is relatively
large the onset of the $I$-resonance will take place early on. As can
be inferred from Fig.~\ref{fig:scheme} that implies that $\bar\nu_e$
will escape the SN as $\bar\nu_2$. For smaller values, though, it may
happen that the $I$-resonance becomes effective only after a few
seconds. This means that during the first seconds of the neutrino
signal $\bar\nu_e$ would leave the star as $\bar\nu_1$ (cases $A$ and
$C$). Then, after some point, the electron fraction would be low enough
to switch on the $I$-resonance, and consequently $\bar\nu_e$ would
enter the Earth as $\bar\nu_2$. This would result in a transition in
the electron antineutrino survival probability from $\bar P_{\rm
  surv}\approx \cos^2\theta_{12}=0.7$ to $\sin^2\theta_{12}=0.3$.
Given the expected hierarchy in the average neutrino energies $\langle
E_{\bar\nu_e} \rangle \lesssim \langle E_{\bar\nu_{\mu,\tau}}
\rangle$, it follows that the change in $Y_e$ would lead to a
hardening of the observed positron spectrum. The effect is quantified
in Fig.~\ref{fig:new2} for different values of $\varepsilon_{\tau\tau}$.
The figure shows the average energy of the $\bar\nu p\to ne^+$ events
for the case of a Megaton water Cherenkov detector exactly as in
Fig.~\ref{fig:shockwave}, but for scenarios $AI$ and $CIa$. One can
see how for $\varepsilon_{\tau\tau}=0.07$ the $I$-resonance condition
is always fulfilled and therefore there is no time dependence. However
for smaller values one can see a rise at a certain moment which
depends on the magnitude of $\varepsilon_{\tau\tau}$.
A similar effect would occur in case $CIb$.

%%%%%%%%%%%%%%%%%%%%%%%%%%%%%%%%%%%%%%%%%%%%%%%%%%%%%%%%%%%%%%%%%%%%%%
\subsection{Earth matter effects}
\label{subsec:eartheffects}
%%%%%%%%%%%%%%%%%%%%%%%%%%%%%%%%%%%%%%%%%%%%%%%%%%%%%%%%%%%%%%%%%%%%%%

Before the shock wave reaches the $H$-resonance layer the dependence
of the neutrino survival probability in the cases we are considering,
on the neutrino energy $E$ is very weak.  However, if neutrinos cross
the Earth before reaching the detector, the conversion probabilities
may become energy-dependent, inducing modulations in the neutrino
energy spectrum. These modulations may be observed in the form of
local peaks and valleys in the spectrum of the event rate $\sigma
F_{\bar\nu_e}^D$ plotted as a function of $1/E$.
These modulations arise in the antineutrino channel only when
$\bar\nu_e$ leave the SN as $\bar\nu_1$ or $\bar\nu_2$.  In the
absence of NSI this happens in cases $A$ and $C$, where $\bar\nu_e$
leave the star as $\bar\nu_1$. In the presence of NSI $\bar\nu_e$ will
arrive at the Earth as $\bar\nu_1$ in cases $BI$, and as $\bar\nu_2$
in case $AI$ and $CIa$.  Therefore its observation would exclude cases
$B$ and $CIb$.
This distortion in the spectra could be measured by comparing the
neutrino signal at two or more different detectors such that the
neutrinos travel different distances through the Earth before reaching
them~\cite{Lunardini:2001pb,Dighe:2003be}.
However these Earth matter effects can be also identified in a single
detector~\cite{Dighe:2003jg,Dighe:2003vm}. 

By analyzing the power spectrum of the detected neutrino events one
can identify the presence of peaks located at the frequencies
characterizing the modulation.
These do not dependend on the primary neutrino spectra, and can be
determined to a good accuracy from the knowledge of the solar
oscillation parameters, the Earth matter density, and the position of
the SN in the sky~\cite{Dighe:2003vm}.  The latter can be determined
with sufficient precision even if the SN is optically obscured using
the pointing capability of water Cherenkov neutrino
detectors~\cite{Tomas:2003xn}.

This method turns out to be powerful in detecting the modulations in
the spectra due to Earth matter effects, and thus in ruling out cases
$B$ and $CIb$. However, the position of the peaks does not depend on
how $\bar\nu_e$ enters the Earth, as $\bar\nu_1$ or $\bar\nu_2$. Hence
it is not useful to discriminate case $AI$ and $CIa$ from the cases
$A$, $C$, and $BI$.

The time dependence of $Y_e$, however, can transform case $B$ into
$BI$, and $C$ with inverse hierarchy into $CIb$, leading respectively
to an appearance and disappearance of these Earth matter effects.
In case $BI$ the presence of the shock wave modulation can spoil a
clear identification of the Earth matter effects. Nevertheless, the
disappearance of the Earth matter effects in the transition from case
$C$ to $CIb$ allows us to pin down case $CIb$.

%%%%%%%%%%%%%%%%%%%%%%%%%%%%%%%%%%%%%%%%%%%%%%%%%%%%%%%%%%%%%%%%%%%%%%
\subsection{Neutronization burst}
\label{subsec:nueburst}
%%%%%%%%%%%%%%%%%%%%%%%%%%%%%%%%%%%%%%%%%%%%%%%%%%%%%%%%%%%%%%%%%%%%%%

The prompt neutronization burst takes place during the first $\sim$
25~ms after the core bounce with a typical full width half maximum of
5--7$\,$ms and a peak luminosity of 3.3--3.5$\times
10^{53}\,$erg$\,$s$^{-1}$. The striking similarity of the neutrino
emission characteristics despite the variability in the properties of
the pre-collapse cores is caused by a regulation mechanism between
electron number fraction and target abundances for electron capture.
This effectively establishes similar electron fractions in the inner
core during collapse, leading to a convergence of the structure of the
central part of the collapsing cores, with only small differences in
the evolution of different progenitors until shock
breakout~\cite{Takahashi:2003rn,Kachelriess:2004ds}.

Taking into account that the SN will be likely to be obscured by dust
and a good estimation of the distance will not be possible, the time
structure of the detected neutrino signal should be used as signature
for the neutronization burst. In Ref.~\cite{Kachelriess:2004ds} it was
shown that such a time structure can be in principle cleanly seen in
the case of a Megaton water Cherenkov detector.
It was also shown how the time evolution of the signal depends
strongly on the neutrino mixing scheme. In the absence of NSI the
$\nu_e$ peak could be observed provided that the $\nu_e$ survival
probability $P_{\nu_e\nu_e}$ is not zero.  As can be seen in
Table~\ref{table:nuschemes} this happens for cases $B$ and $C$.
However for case $A$ (normal mass hierarchy and ``large''
$\theta_{13}$), $\nu_e$ leaves the SN as $\nu_3$. This leads to a
survival probability $P_{\nu_e\nu_e}\approx \sin^2\theta_{13}\lesssim
10^{-1}$, and therefore the peak remains hidden.

Let us now consider the situation where NSI are prensent.  For normal
mass hierarchy $\nu_e$, which is born as $\nu_2^m$ passes through
three different resonances, $I,~H$ and $L$. Whereas $I$ and $L$ will
be adiabatic, the fate of $H$ will depend on the value of
$\theta_{13}$.  For ``large'' values, case $AI$, the $H$-resonance
will also be adiabatic.  This implies that $\nu_e$'s will leave as
$\nu_2$, the survival probability will be $P_{\nu_e\nu_e}\approx
\sin^2\theta_{12} \approx 0.3$, and therefore the peak will be seen,
as in cases $B$ and $C$. If $\theta_{13}$ happens to be very small, case
$CIa$, then $H$ will be strongly non-adiabatic and therefore $\nu_e$
will leave the star as $\nu_3$. As a consequence the neutronization
peak will not be seen.

\begin{figure}
  \begin{center}
    \includegraphics[width=0.45\textwidth]{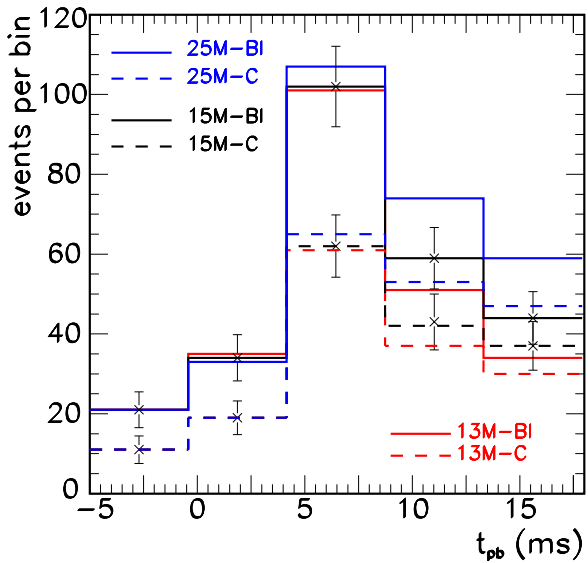}
  \end{center}
  \caption{Number of events from the elastic scattering on electrons,
    per time bin in a Megaton water Cherenkov detector for a SN at 10
    kpc for cases $C$ (dashed lines) and $BI$ (solid lines). Different
    progenitor masses have been assumed: 13~$M_\odot$ (n13) in red,
    15~$M_\odot$ (s15s7b2) in black, and 25~$M_\odot$ (s25a28) in
    blue. 1-sigma errors are also shown for the 15~$M_\odot$ case.}
  \label{fig:nuepeak}
\end{figure}

For inverse mass hierarchy, $\nu_e$ is born as $\nu_1^m$ and traverses
adiabatically $I$ and $L$. This implies that they will leave the star
as $\nu_1$ and therefore the peak will also be observed. However now
the survival probability will be larger, $P_{\nu_e\nu_e}\approx
\cos^2\theta_{12} \approx 0.7$. Thus for a given known normalization,
i.e. the distance to the SN, one expects a larger number of events
during the neutronization peak in this case. In Fig.~\ref{fig:nuepeak}
we show the expected number of events per time bin in a water
Cherenkov detector in the case of a SN exploding at 10 kpc, for two
different neutrino schemes, $C$ and $BI$, and for different SN
progenitor masses. One can see how the difference due to the larger
survival probability is bigger than the typical error bars, associated
to the lack of knowledge of the progenitor mass.

Two comments are in order. The neutronization $\nu_e$ burst takes
place during the first milliseconds, before strong deleptonization
takes place. As a result, in contrast to other observables we have
considered in this paper, here the $I$-resonance will only occur for
$\varepsilon^{I}\gtrsim 10^{-1}$.
On the other hand in the presence of additional NSI with electrons
this would significantly affect the $\nu-e$ cross sections, and
consequently the results presented here.

%%%%%%%%%%%%%%%%%%%%%%%%%%%%%%%%%%%%%%%%%%%%%%%%%%%%%%%%%%%%%%%%%%%%%%
\section{Summary}
\label{sec:summary}
%%%%%%%%%%%%%%%%%%%%%%%%%%%%%%%%%%%%%%%%%%%%%%%%%%%%%%%%%%%%%%%%%%%%%%

We have analyzed the possibility of observing clear signatures of
non-standard neutrino interactions from the detection of neutrinos
produced in a future galactic supernova.

In Secs.~\ref{sec:neutrino-evolution} and~\ref{sec:two-regimes} we
have re-considered effect of $\nu-d$ non-standard interactions on the
neutrino propagation through the SN envelope within a three-neutrino
framework.  In contrast to previous works we have analyzed the
neutrino evolution in both the more deleptonized inner layers and the
outer regions of the SN envelope. We have also taken into account the
time dependence of the SN density and electron fraction profiles.

First we have found that the small values of the electron fraction
typical of the former allows for internal NSI-induced resonant
conversions, in addition to the standard MSW-H and MSW-L resonances of
the outer envelope. These new flavor conversions take place for a
relatively large range of NSI parameters, namely
$|\varepsilon_{\alpha\alpha}|$ between $10^{-2}-10^{-1}$, and
$|\varepsilon_{e\tau}|\gtrsim {\rm few}\times 10^{-5}$, currently
allowed by experiment.  For this range of strengths, in particular
$\varepsilon_{\tau\tau}$, non-standard interactions can significantly
affect the adiabaticity of the $H$-resonance.  On the other hand the
NSI-induced resonant conversions may also lead to the modulation of
the $\bar\nu_e$ spectra as a result of the time dependence of the
electron fraction.

In Sec.~\ref{sec:observables} we have studied the possibility of
detecting NSI effects in a Megaton water Cherenkov detector using the
modulation effects in the $\bar\nu_e$ spectrum due to (i) the passage
of shock waves through the SN envelope, (ii) the time dependence of
the electron fraction and (iii) the Earth matter effects; and, finally,
through the possible detectability of the neutronization $\nu_e$
burst.
Note that observable (ii) turns out to be complementary to the
observation of the shock wave passage, (i), and offers the possibility
to probe NSI effects also for normal hierarchy neutrino spectra.

In Table~\ref{table:summary} we summarize the results obtained for
different neutrino schemes.
We have found that observable (i) can clearly indicate the existence
of NSI in the case of inverse mass hierarchy and large $\theta_{13}$
(case $BI$).
On the other hand, observable (ii) allows for an identification of NSI
effects in the other cases, normal mass hierarchy (cases $AI$ and
$CIa$) and inverse mass hierarchy and small $\theta_{13}$ (case
$CIb$). 
Therefore a positive signal of either observable (i) or (ii) would
establish the existence of NSI. In the latter case this would,
however, leave a degeneracy among cases $AI$, $CIa$, and $CIb$.
Such degeneracy can be broken with the help of observables (iii) and
the observation of the neutronization $\nu_e$ burst.
The detection of Earth matter effects during the whole supernova
neutrino signal would rule out case $CIb$ since, as discussed in
Sec.~\ref{subsec:eartheffects}, a disappearance of Earth matter
effects would take place due to a transition from $C$ to $CIb$.
Finally, the (non) observation of the neutronization burst can be used
to distinguish between cases $AI$ and $CIa$.

Similarly, other degeneracies in Table \ref{table:summary} may be
lifted by suitably combining different observables. For example, a
negative of observable (ii) could mean either negligible NSI strengths
or (NU) NSI parameter values so large that the internal resonance is
always present. In this case one could use the observation of the
neutronization burst in order to establish the presence of NSI for the
case of inverse mass hierarchy. In addition the observation of the
shock wave imprint in the $\bar\nu_e$ spectrum would provide
additional information on $\theta_{13}$.

%%%%

In conclusion, by suitably combining all observables one may establish
not only the presence of NSI, but also the mass hierarchy and probe
the magnitude of $\theta_{13}$.

\begin{table}
\begin{center}
\begin{tabular}{|c|c|c|c|c|c|c|c|}
\hline
 Scheme & Hierarchy & $\sin^2\theta_{13}$ & NSI  & shock & $Y_e$ &
 Earth & $\nu_e$ burst  \\
\hline
\hline
$A$ & normal & $\gtrsim 10^{-4}$ & No  &
 No & No & Yes & No  \\
\hline
$B$ & inverted & $\gtrsim 10^{-4}$ & No  & 
 Yes & No & No & Yes  \\ 
\hline
$C$ & any & $\lesssim 10^{-6}$ & No &  No &  No & Yes & Yes \\
\hline
\hline
$AI$ & normal & $\gtrsim 10^{-4}$ & Yes  & 
No & Yes & Yes & Yes \\
\hline
$BI$ & inverted & $\gtrsim 10^{-4}$ & Yes &  
Yes$^\star$ & No & Yes & Yes$^\star$  \\  
\hline
$CIa$ & normal & $\lesssim 10^{-6}$ & Yes  & 
No & Yes  & Yes & No  \\
\hline
$CIb$ & inverted & $\lesssim 10^{-6}$ & Yes & 
No & Yes & No & Yes$^\star$ \\
\hline
\end{tabular}
\end{center}
\caption{Expectations for the observables discussed in the text:
  modulation of the $\bar\nu_e$ spectrum due to the shock wave
  passage, the time variation of $Y_e$, the Earth effect, and the
  observation of the $\nu_e$ burst within various neutrino
  schemes. Asterisks indicate that the effect differs from that
  expected in the absence of NSI. See text.  }
\label{table:summary}
\end{table}

%%%%%%%%%%%%%%%%%%%%%%%%%%%%%%%%%%%%%%%%%%%%%%%%%%%%%%%%%%%%%%%%%%%%%%
%% Acknowledgments %%%%%%%%%%%%%%%%%%%%%%%%%%%%%%%%%%%%%%%%%%%%%%%%%%%
%%%%%%%%%%%%%%%%%%%%%%%%%%%%%%%%%%%%%%%%%%%%%%%%%%%%%%%%%%%%%%%%%%%%%%
\section*{Acknowledgments}

The authors wish to thank H-Th. Janka, O.~Miranda, S.~Pastor,
Th.~Schwetz, and M.~T\'ortola for fruitful discussions. Work supported
by the Spanish grant FPA2005-01269 and European Network of Theoretical
Astroparticle Physics ILIAS/N6 under contract number
RII3-CT-2004-506222. A. E. has been supported by a FPU grant from the
Spanish Government.  R. T. has been supported by the Juan de la Cierva
program from the Spanish Government and by an ERG from the European
Commission.

%%%%%%%%%%%%%%%%%%%%%%%%%%%%%%%%%%%%%%%%%%%%%%%%%%%%%%%%%%%%%%%%%%%%%%%
\section*{References}
%%%%%%%%%%%%%%%%%%%%%%%%%%%%%%%%%%%%%%%%%%%%%%%%%%%%%%%%%%%%%%%%%%%%%%%

\def\baselinestretch{1.2}
% \bibliographystyle{h-physrev4} 
% \bibliography{valle-ref,snova,nu-rev06,parke-ref}

\end{document}